\documentclass[useAMS,usenatbib]{mn2e}
\usepackage{times}
\usepackage{graphicx}

\usepackage{amsmath}
\usepackage{amssymb}

\usepackage[T1]{fontenc}
\usepackage{aecompl}
\usepackage{epstopdf}

\begin{document}

\title[Limits on minihalo star formation]{Limits on Population III star formation in minihaloes implied by \emph{Planck}}

\author[E. Visbal et al.]{Eli Visbal$^1$\thanks{visbal@astro.columbia.edu} \thanks{Columbia Prize Postdoctoral Fellow in the Natural Sciences}, Zolt\'{a}n Haiman$^1$, Greg L. Bryan$^1$ \\ $^1$Department of Astronomy, Columbia University, 550 West 120th Street, New York, NY, 10027, U.S.A. }

\maketitle

\begin{abstract}
Recently, \emph{Planck} measured a value of the cosmic microwave background (CMB) optical depth due to electron scattering of $\tau=0.066 \pm 0.016$. Here we show that this low value leaves essentially no room for an early partial reionisation of the intergalactic medium (IGM) by high-redshift Population III (Pop III) stars, expected to have formed in low-mass minihaloes. We perform semi-analytic calculations of reionisation which include the contribution from Pop II stars in atomic cooling haloes, calibrated with high-redshift galaxy observations, and Pop III stars in minihaloes with feedback due to Lyman-Werner (LW) radiation and metal enrichment.  We find that without LW feedback or prompt metal enrichment (and assuming a minihalo escape fraction of 0.5) the Pop III star formation efficiency cannot exceed $\sim{\rm a~few}\times 10^{-4}$, without violating the constraints set by \emph{Planck} data. This excludes massive Pop III star formation in typical $10^6 M_\odot$ minihaloes. Including LW feedback and metal enrichment alleviates this tension, allowing large Pop III stars to form early on before they are quenched by feedback. We find that the total density of Pop III stars formed across cosmic time is $\lesssim 10^{4-5}~M_\odot~{\rm Mpc^{-3}}$ and does not depend strongly on the feedback prescription adopted. Additionally, we perform a simple estimate of the possible impact on reionisation of X-rays produced by accretion onto black hole remnants of Pop III stars. We find that unless the accretion duty cycle is very low ($\lesssim 0.01$), this could lead to an optical depth inconsistent with \emph{Planck}.
\end{abstract}


\section{Introduction}
Understanding cosmic reionisation is currently one of the most exciting frontiers in astrophysical cosmology \citep{2013fgu..book.....L}. Constraints from Ly$\alpha$ forest measurements indicate that the IGM was completely ionised around $z\approx 6-7$ \citep[][]{2006ARA&A..44..415F, 2011Natur.474..616M, 2013MNRAS.428.3058S, 2015MNRAS.447..499M}, however the exact ionisation history at higher redshifts remains unclear. The optical depth due to electron scattering of the CMB, $\tau$, provides an important constraint, however because it is only one number (an integral constraint on the ionisation evolution), it is degenerate with different ionisation histories. 
Future data, including radio observations of 21cm emission from neutral hydrogen in the IGM \citep{2006PhR...433..181F, 2012RPPh...75h6901P}, will give a more detailed picture of the reionisation process. Comparing this data with theory will provide an important test of the standard model of cosmology and yield information about ionising sources which may be too faint to observe directly.

Reionisation is thought to be primarily driven by UV photons from stars. The first stars in the Universe are expected to form from metal-free gas in $\sim 10^6 ~M_\odot$ dark matter ``minihaloes'' \citep{1996ApJ...464..523H, 1997ApJ...474....1T, 2002Sci...295...93A,BCL02}.  These so-called Population III (Pop III) stars are predicted to be more efficient at producing ionising radiation than metal-enriched (Pop I/II) stars \citep{2000ApJ...528L..65T, 2002A&A...382...28S, 2003A&A...397..527S}. This ionising efficiency would be further enhanced if Pop III stars form with a top-heavy initial mass function (IMF). Cosmological hydrodynamical simulations have been utilised to study the formation of Pop III stars \citep[e.g.][]{2010MNRAS.403...45S,2011ApJ...727..110C,2012MNRAS.424..399G,2014ApJ...781...60H,2013RPPh...76k2901B,2015ComAC...2....3G}, but the IMF remains highly uncertain.

Formation of Pop III stars in minihaloes requires efficient molecular cooling. As stars form, a background of Lyman-Werner (LW) radiation builds up over cosmic time. Eventually this LW radiation is strong enough to photo-dissociate molecular hydrogen, inhibiting additional star formation in minihaloes \citep{1997ApJ...476..458H, 2001ApJ...548..509M, 2007ApJ...671.1559W, 2008ApJ...673...14O, 2011MNRAS.418..838W, 2014MNRAS.445..107V}. At this point, only haloes with virial temperatures $T_{\rm vir} \gtrsim 10^4~{\rm K}$ (hereafter ``atomic cooling haloes'') can form stars through atomic hydrogen cooling. Pop III stars could form in these haloes if they contain pristine gas, but once they become metal-enriched from previous generations of stars, Pop I/II star formation occurs.

Recently \emph{Planck} has reported an improved measurement of the CMB electron scattering optical depth, $\tau = 0.066 \pm 0.016$ \citep{2015arXiv150201589P}. This is lower than previous measurements from \emph{WMAP} \citep{2011ApJS..192...18K} and leaves less room for an early partial reionisation. In this paper, we examine the constraints this new measurement puts on the production of Pop III stars in minihaloes \citep[see][for related studies based on the \emph{WMAP} cosmological parameters]{2006ApJ...650....7H,2006MNRAS.373..128G}. We utilise a semi-analytic model that includes the contribution from both atomic cooling haloes, calibrated with observations of $z \approx 6$ galaxies, and minihaloes with a self-consistent treatment of LW feedback. Using a simple treatment, we also include the possible effects of metal enrichment in minihaloes from Pop III supernovae, which causes a transition to Pop II star formation. 

We compute the ionisation history and corresponding CMB optical depth for a variety of model parameterisations with and without LW feedback and metal enrichment and find that without LW feedback or metal enrichment, massive Pop III stars cannot form efficiently in minihaloes without violating the \emph{Planck} constraints. When LW feedback and metal enrichment are included, massive Pop III stars could form efficiently early on, but they are suppressed at lower redshifts, reducing the optical depth sufficiently to be consistent with \emph{Planck}. We also find that, irrespective of the feedback prescription used, the total density of Pop III stars formed over all cosmic time cannot exceed $\approx 10^{4-5}~M_\odot~{\rm Mpc^{-3}}$ without violating the \emph{Planck} optical depth constraints.
 
Note that \cite{2015ApJ...802L..19R} recently performed an analysis of reionisation and the \emph{Planck} optical depth. Their study is empirical, focusing on the ionisation history implied by the observed UV luminosity function (LF) (with a modest extrapolation to fainter galaxies), while we address the implications for lower-mass minihaloes and Pop III stars. Similarly, \cite{2015arXiv150505507M} use a semi-analytic model to show that reionisation from Pop II star formation alone (i.e. no Pop III) is consistent with the \emph{Planck} optical depth measurement and high-redshift quasar absorption spectra. They do not attempt the put upper limits on the amount of Pop III star formation permitted, which is the primary goal of this work. 

In addition to the UV photons from stars, X-rays from black hole accretion could potentially contribute to reionisation \citep{2001ApJ...563....1V, 2004ApJ...604..484M, 2004MNRAS.352..547R, 2009ApJ...703.2113V}. We perform a simple calculation to estimate how much gas the black hole remnants of massive Pop III stars could accrete with a high radiative efficiency without producing a $\tau$ inconsistent with the \emph{Planck} measurement. We find that for our fiducial model with LW feedback and moderate minihalo star formation efficiency, black hole remnants of Pop III stars cannot accrete at the Eddington limit with a duty cycle higher than $\sim 0.01$. This suggests that either massive Pop III stars are uncommon or there is some feedback mechanism which prevents sustained accretion \citep[e.g.][]{Alvarez+09,2012MNRAS.425.2974T}.

This paper is structured as follows. In \S~2, we describe our reionisation model including the self-consistent prescription for LW feedback, the various model parameters and their chosen fiducial values, and our simple treatment of metal enrichment from Pop III supernovae. We present the results of this model in \S~3. In \S~4, we perform a simple calculation to estimate how the growth of black hole remnants of massive Pop III stars would impact reionisation. Finally, we discuss our results and conclusions in \S~5. Throughout we assume a $\Lambda$CDM cosmology consistent with the latest constraints from \emph{Planck} \citep{2014A&A...571A..16P}: $\Omega_\Lambda=0.68$, $\Omega_{\rm m}=0.32$, $\Omega_{\rm b}=0.049$, $h=0.67$, $\sigma_8=0.83$, and $n_{\rm s} = 0.96$.

\section{Reionisation model}
Here we outline our semi-analytic model of reionisation. In \S~2.1 and \S~2.2, we explain how we compute the ionisation history and describe our self-consistent treatment of LW feedback. In \S~2.3 we discuss the physical parameters of the model and their fiducial values. Finally, in \S~2.4 we introduce our simple treatment of metal enrichment due to Pop III supernovae.

\vspace{-\baselineskip}
\subsection{Ionised filling factor}
We model the global reionisation process by considering star formation in dark matter haloes, closely following \citet{2003ApJ...595....1H}. An ionising efficiency and associated ionised volume are assigned to each dark matter halo and the total halo abundance is computed analytically with the Sheth-Tormen mass function \citep{1999MNRAS.308..119S}. We assign different ionising efficiencies to minihaloes which we assume host Pop III stars and atomic cooing haloes hosting Pop II stars.  We also assume that in regions of the IGM that have already been ionised, the increased Jeans mass of the photo-heated gas prevents star formation below a characteristic halo mass, $M_{\rm i}$. The other important mass scales are the minimum minihalo mass and the atomic cooling mass, which are denoted by $M_{\rm m}$ and $M_{\rm a}$ (we discuss the fiducial values of these masses below).  It follows that the total ionised filling factor as a function of redshift, $Q(z)$, is given by
\begin{multline}
\label{fill_fac}
Q(z) = \rho_{\rm b}(z) \int_\infty^z dz' \biggl [ \epsilon_{\rm a} \frac{dF_{\rm coll, i}}{dz}(z') + \\  \left (1-Q(z') \right )\times \left ( \epsilon_{\rm a} \frac{dF_{\rm coll, a}}{dz}(z') +  \epsilon_{\rm m} \frac{dF_{\rm coll, m}}{dz}(z')   \right) \biggr]  \tilde{V}(z',z),
\end{multline}
where $\rho_{\rm b}(z)$ is the mean cosmic baryon density. The ionising efficiency in minihaloes (i.e. number of ionising photons escaping into the IGM per baryon incorporated into a dark matter halo) is given by  $\epsilon_{\rm m} = f_{\rm *, m }f_{\rm esc, m}\eta_{\rm ion, m}$, where the star formation efficiency, $f_{\rm *, m}$, is the fraction of baryons in minihaloes that form stars, $f_{\rm esc, m}$ is the ionising photon escape fraction, and $\eta_{\rm ion, m}$ is the number of ionising photons produced per baryon incorporated into stars. Similarly, the ionising efficiency above the atomic cooling mass is $\epsilon_{\rm a} = f_{\rm *, a}f_{\rm esc, a}\eta_{\rm ion, a}$. 
The cosmic mass fraction collapsed into dark matter haloes above the ionised IGM feedback threshold is given by 
\begin{equation}
F_{\rm coll, i}(z) = \frac{1}{\Omega_{\rm m} \rho_{\rm c}} \int_{M_{\rm i}}^\infty dM M \frac{dn}{dM}(z),
\end{equation}
where $\rho_{\rm c}$ is the critical cosmological density and $\frac{dn}{dM}$ is the Sheth-Tormen mass function. 
Similarly, the collapsed fractions for atomic cooling haloes below $M_{\rm i}$ and for minihaloes are
\begin{equation}
\label{atom_frac}
F_{\rm coll, a}(z) = \frac{1}{\Omega_{\rm m} \rho_{\rm c}} \int_{M_{\rm a}}^{M_{\rm i}} dM M \frac{dn}{dM}(z),
\end{equation} 

\begin{equation}
\label{minifrac}
F_{\rm coll, m}(z) = \frac{1}{\Omega_{\rm m} \rho_{\rm c}} \int_{M_{\rm m}}^{M_{\rm a}} dM M \frac{dn}{dM}(z).
\end{equation}

We denote the ionised volume of an HII region per unit gas mass in a dark matter halo and unit ionising efficiency at redshift $z$ as $\tilde{V}(z_{\rm on},z)$, where $z_{\rm on}$ is the redshift corresponding to the formation of the halo (the total ionising volume for each halo is given by $V = \epsilon  \frac{\Omega_{\rm b}}{\Omega_{\rm m}} M \tilde{V}$). To determine this value we solve the equation of motion of the ionisation front, $R_{\rm i} =  \left ( V \frac{3}{4 \pi} \right )^{1/3}$,
\begin{equation}
\frac{dR^3_i}{dt} = 3H(z)R^3_{\rm i} + \frac{3\dot{N}_\gamma}{4\pi \langle n_H \rangle } - C(z) \langle n_H \rangle \alpha_{\rm B} R^3_{\rm i},
\end{equation}
where $H(z)$ is the Hubble parameter, $\alpha_{\rm B}= 2.6\times 10^{-13} {\rm cm^3s^{-1}}$ is the case~B recombination coefficient of hydrogen at $T=10^4$ K, $\langle n_H \rangle$ is the mean cosmic hydrogen density, and $C(z) \equiv \langle n_{\rm HII}^2 \rangle / \langle n_{\rm HII} \rangle^2 $ is the clumping factor of the ionised IGM. For each $M_\odot$ of star forming gas with ionising efficiency normalised to unity, we assume the time-dependent rate of ionising photon emission, $\dot{N}_\gamma$, is
\begin{eqnarray}
\dot{N}_\gamma = \begin{cases}  \dot{N}_0 & (t \le 10^{6.5} \rm{yr}), \\ \dot{N}_0\times(t/10^{6.5} \rm{yr})^{-4.5} & (t > 10^{6.5} \rm{yr}),  \end{cases}
\end{eqnarray}
where $\dot{N}_0 = 9.25\times 10^{42} {\rm s^{-1}}$. Over the entire lifetime of the stellar population this normalisation yields 1 ionising photon per baryon incorporated into stars. The exact form of $\dot{N}_\gamma(t)$ does not impact our results since the majority of the photons are produced much faster than the Hubble time for all redshifts relevant to our calculations.

\begin{table*}
\centering
\caption{\label{table} Physical parameters used in our reionisation model and their fiducial values. See \S~2.3 for more details. }
\begin{tabular}{c l| c || c }
Parameter & Description & Fiducial Value  \\
\hline
$C(z)$ & clumping parameter & $ 2 \left ( \frac{1+z}{7}  \right )^{-2} + 1$   \\ [1ex]
 $\epsilon_{\rm a}$ & ionising efficiency of atomic cooling haloes  & $ 600 f_{\rm *, a}$   [ $400 f_{\rm *, a}$ in redshift-independent model] \\ [1ex]
 $ f_{\rm *, a}$ &   atomic cooling halo star formation efficiency &see Figure \ref{f_star_vs_z}  [0.05 in redshift-independent model]  \\ [1ex]
 $M_{\rm a}$  & atomic cooling mass &$5.4 \times 10^7 \left(\frac{1+z}{11} \right )^{-1.5} M_\odot$  \\  [1ex]
 $M_{\rm i}$ & ionised IGM feedback mass & $1.5 \times 10^8 \left(\frac{1+z}{11} \right )^{-1.5} M_\odot$   \\   [1ex] 
 $\eta_{\rm LW, a}$ & LW photons per stellar baryon in atomic cooling haloes &  4000   \\   [1ex] 
$\epsilon_{\rm m}$ & ionising efficiency of minihaloes & $ 40000 f_{\rm *, m}$ \\ [1ex]
$f_{\rm *, m}$ & minihalo star formation efficiency & treated as free parameter \\   [1ex] 
 $M_{\rm m}$ & minimum minihalo mass for star formation via ${\rm H_2}$ & $ 2.5 \times 10^5 \left (  \frac{1+z}{26}  \right )^{-1.5} \left ( 1 + 6.96 \left (4\pi J_{\rm LW}(z) \right)^{0.47}\right ) M_\odot$   \\   [1ex]  
 $\eta_{\rm LW, m}$ & LW photons per stellar baryon in minihaloes &  80000   \\   [1ex]  
\end{tabular}
\end{table*}

\subsection{Lyman-Werner feedback}
As described above, the LW background can dissociate molecular hydrogen, increasing the minimum mass of minihaloes that host Pop III star formation. We incorporate this into our model by self-consistently computing the mean LW background, $J_{\rm LW}(z)$, and using it to set the minimum mass of minihaloes hosting Pop III star formation, $M_{\rm m}$ \citep[see, e.g.][]{2000ApJ...534...11H}. We assume the minimum mass is equal to 
\begin{equation}
\label{mini_mass}
M_{\rm m} =  2.5 \times 10^5 \left (  \frac{1+z}{26}  \right )^{-1.5} \left ( 1 + 6.96 \left (4\pi J_{\rm LW}(z) \right)^{0.47}\right ), 
\end{equation}
where $J_{\rm LW}$ is in units of $10^{-21}~{\rm ergs~s^{-1}~cm^{-2}~Hz^{-1}~Sr^{-1}}$ \citep{2013MNRAS.432.2909F}. This formula gives an increase in minimum mass due to LW radiation that is consistent with the simulations of \citet{2001ApJ...548..509M},  \citet{2008ApJ...673...14O}, and  \citet{2007ApJ...671.1559W}. The mass for  $J_{\rm LW}=0$ is taken as the ``optimal fit'' from \cite{2012MNRAS.424.1335F} which was calibrated with the simulations of \citet{2011MNRAS.413..543S} and \citet{2011ApJ...736..147G}.

We compute $J_{\rm LW}$ by making a simple ``screening'' assumption that the IGM is nearly transparent to LW photons until they are redshifted into a Lyman series line and absorbed, removing them from the LW band. We approximate this by assuming that at a redshift of $z$ all emitted LW photons can be seen from sources out to $z_{\rm max}= 1.015 \times z$.  The factor of 1.015 is used because 1.5 per cent is approximately the amount a typical LW photon can be redshifted before reaching a Lyman series line. In reality, the LW attenuation as a function of frequency will follow a more complicated characteristic ``sawtooth'' shape. However, we expect our simple approximation to be reasonably accurate. We find that in our fiducial model with constant $f_{\rm *,a}$ described below, our screening reduces the LW background by roughly an order of magnitude at $z\approx 15$ and a factor of a few at $z\approx30$, which is consistent with more sophisticated treatments \cite[e.g.][]{2000ApJ...534...11H,Ricotti+01,2009ApJ...695.1430A,2011MNRAS.418..838W}.

Given these assumptions the LW background is
\begin{equation}
J_{\rm LW}(z) = \frac{c(1+z)^3}{4 \pi} \int^z_{z_{\rm max}} dz' \frac{dt_{\rm H}}{dz'} \epsilon_{\rm LW}(z'),
\end{equation}
where $c$ is the speed of light and $t_{\rm H}$ is the Hubble time.
The LW luminosity per frequency per comoving volume is given by 
\begin{equation}
\epsilon_{\rm LW}(z) =  \left ( \frac{\rm SFRD_{\rm a}}{m_{\rm p}} \eta_{\rm LW, a} + \frac{\rm SFRD_{\rm m}}{m_{\rm p}} \eta_{\rm LW, m}  \right)E_{\rm LW} \Delta \nu_{\rm LW}^{-1},
\end{equation}
where SFRD is the star formation rate density, $m_{\rm p}$ is the proton mass, $\eta_{\rm LW}$ is the number of LW photons per baryon produced in stars, $E_{\rm LW}=1.9 \times 10^{-11}~{\rm erg}$ and $\Delta \nu_{\rm LW}=5.8 \times 10^{11}~ \rm{Hz}$. The subscripts ``a'' and ``m'' denote atomic cooling haloes and minihaloes as above. We compute ${\rm SFRD}_{\rm a}$ from the collapsed fractions of haloes above the atomic cooling threshold,
\begin{equation}
{\rm SFRD}_{\rm a}(z) = \rho_{\rm b} f_{\rm *, a} \frac{dF_{\rm coll, a}}{dt} \left[1-Q(z) \right] +  \rho_{\rm b} f_{\rm *, a} \frac{dF_{\rm coll, i}}{dt}.
\end{equation}
Similarly, the minihalo SFRD is given by 
\begin{equation}
\label{mini_sfrd}
{\rm SFRD}_{\rm m}(z) = \rho_{\rm b} f_{\rm *, m} \frac{dF_{\rm coll, m}}{dt} \left(1-Q(z) \right).
\end{equation}
 Note that throughout this work, we assume a LW photon escape fraction of unity. This is consistent with the results of \cite{2015arXiv150604796S}, who find that for a single large Pop III star in a moderate size minihalo ($6.9\times 10^5~ M_\odot$ and $2.1 \times 10^6~ M_\odot$ ) the ``far-field''  LW escape fraction is unity. For a single star in a halo closer to the atomic cooling threshold ($1.2 \times 10^7~ M_\odot$), they find an escape fraction of $\approx 0.7$. Due to the 1D nature of these calculations, and since we expect a higher rate of star-formation in halos close to the atomic cooling threshold, these values should be taken as lower limits on the escape fraction.

To determine the form of $J_{\rm LW}(z)$ self-consistently, we iteratively compute the entire star formation evolution, reionisation history, and $J_{\rm LW}(z)$, using the $J_{\rm LW}(z)$ computed in previous steps until we achieve convergence.

\subsection{Model parameters}
Here we discuss the physical parameters that enter our calculations and their fiducial values. Without including LW feedback these parameters are $C(z)$, $\epsilon_{\rm a}$, $\epsilon_{\rm m}$, $M_{\rm m}$, $M_{\rm a}$,  and $M_{\rm i}$. With LW feedback $f_{\rm *, a}$, $f_{\rm *, m}$, $\eta_{\rm ion, a}$, $\eta_{\rm ion, m}$, $\eta_{\rm LW, a}$, and $\eta_{\rm LW, m}$ are also required.  As explained below, we consider two different models of the star formation efficiency in atomic cooling haloes. In the ``redshift-dependent" model, $f_{\rm *, a}$ varies as a function of cosmic time and in the ``redshift-independent" model it is constant. We summarize the fiducial parameter choices for both of these models in Table \ref{table}. We treat $f_{\rm *, m}$ as a free parameter and determine how it is constrained by \emph{Planck} in \S~3. To demonstrate that these constraints do not depend strongly on our fiducial parameter choices we vary each of the other parameters subject to the constraint that reionisation is completed by $z=6$ and find that our conclusions remain robust.

Next we describe the choice for each of our fiducial parameters. We adopt a redshift-dependent clumping factor of the ionised IGM parameterised by 
\begin{equation}
C(z) = 2 \left ( \frac{1+z}{7}  \right )^{-2} + 1.
\end{equation} 
 This formula is similar to the clumping factor found in the Illustris simulation for gas below 20 times the mean baryon density \citep{2015arXiv150300734B}.  \cite{2012MNRAS.427.2464F} find a clumping factor of the ionised IGM similar to this relation as well. 

\begin{figure}
\includegraphics[width=88mm]{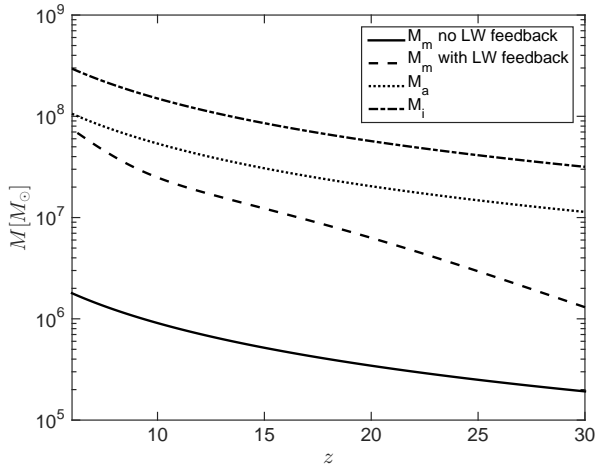}
\caption{\label{mass_scales} Important mass scales in our reionisation model as a function of redshift. For the case with LW feedback, we use the $J_{\rm LW}(z)$ from Figure \ref{J_plot}.  }
\end{figure}

For the atomic cooling mass, we take a fiducial value of $M_{\rm a}=5.4 \times 10^7 \left(\frac{1+z}{11} \right )^{-1.5}$. This corresponds to the typical minimum mass of haloes that are able cool in the absence of ${\rm H_2}$ in the simulations of \cite{2014MNRAS.439.3798F}. We set the fiducial ionised IGM feedback mass to $M_{\rm i}= 1.5 \times 10^8 \left(\frac{1+z}{11} \right )^{-1.5}$. This corresponds to a halo circular velocity of 20 km s$^{-1}$, which \cite{2004ApJ...601..666D} find sets the mass scale where feedback becomes important.  The minimum minihalo mass is assumed to follow Eq. \ref{mini_mass}, which is calibrated with the simulations mentioned above. We plot the important mass scales discussed here in Figure \ref{mass_scales}.

Converting between star formation efficiency and ionisation efficiency requires values for $\eta_{\rm ion}$ and $f_{\rm esc}$ for both atomic cooling haloes and minihaloes.  We assume $\eta_{\rm ion, a}=4000$, which corresponds to a stellar population with a Salpeter IMF from $0.1-100~M_\odot$ and metallicity of $Z=0.0004$ (see Table 1 in \cite{2007MNRAS.377..285S} to see how this quantity changes for different IMFs and metallicities). For minihaloes we adopt $\eta_{\rm ion, m}=80000$, which is accurate for Pop III stars with masses greater than $\sim 200~M_\odot$ \citep{2002A&A...382...28S}. Smaller stars would reduce this value (e.g a reduction by a factor of $\sim$1.40 for 80 $M_\odot$ stars or $\sim3$ for $40~M_\odot$ stars).  For the escape fractions, we take $f_{\rm esc, a}=0.15$ in the redshift-dependent  $f_{\rm *,a}$ model and $f_{\rm esc, a}=0.1$ in the redshift-independent $f_{\rm *,a}$ model. We assume $f_{\rm esc, m}=0.5$ in minihaloes. These values are generally consistent with the simulations of \cite{2014MNRAS.442.2560W}, but we regard this parameter as uncertain. Simulations of minihaloes have often found escape fractions higher than 0.5 \citep{2004ApJ...610...14W,2004ApJ...613..631K,2006ApJ...639..621A}, suggesting that our fiducial choice is conservative (a low assumed value puts weaker limits on Pop III star formation).    As discussed below, we find that $\epsilon_{\rm a} (\propto f_{\rm esc, a})$ cannot be taken to be significantly lower than our fiducial model without reionisation occurring at $z<6$. Higher values would lead to even more stringent constraints on $f_{\rm *, m}$ than we present below. When including LW feedback, we assume one LW photon per ionising photon (i.e. $\eta_{\rm LW, a}=4000$ and $\eta_{\rm LW, m} = 80000$). This is a reasonably good assumption for a wide range of IMFs and metallicities \citep[see table 4 in][]{2002A&A...382...28S}. Note that only the combinations $\eta_{\rm ion,a} f_{\rm esc, a}$,  $\eta_{\rm ion,m} f_{\rm esc, m}$, $f_{\rm *, a}\eta_{\rm LW, a }$, and $f_{\rm *, m}\eta_{\rm LW, m}$ appear in our model. None of the individual parameters appear alone outside of these products.

We calibrate our fiducial values of $\epsilon_{\rm a}$ and $f_{\rm *,a}$ with observations of the UV LF at $z\approx6$ and abundance matching. For a given absolute UV magnitude, $M_{\rm UV}$ (at a rest-frame wavelength of 1600 \AA), we find a corresponding halo mass, $M$, satisfying
\begin{equation}
\int_{-\infty}^{M_{\rm UV}} dM_{\rm UV} \phi(M_{\rm UV}) = \epsilon_{\rm duty} \int_M^\infty dM \frac{dn}{dM},
\end{equation}
where $\phi(M_{\rm UV})$ is the best fit LF function at $z\approx 5.9$ from  \cite{2015ApJ...803...34B} (a Schechter function with $M_{*} = -20.94$, $\phi_* = 0.5 \times 10^{-3}$, and $\alpha = -1.87$). We assume that only a fraction of dark matter haloes host bright UV galaxies at a given time. This is parameterised with the duty cycle, taken to be $\epsilon_{\rm duty}=0.1$ in the fiducial case. This value is consistent with galaxy clustering measurements \citep{2014ApJ...793...17B}. Once we associate an absolute UV magnitude with each halo mass in the relevant range, we convert the magnitude to a SFR with the following relation
\begin{equation}
\frac{\rm SFR}{M_\odot~{\rm yr}^{-1}}= 2.24\times10^{-28} \frac{L_{\nu}}{\rm ergs~s^{-1}~Hz^{-1}}.
\end{equation}
This is the ratio of the dust-corrected SFRD to the luminosity density given in \cite{2015ApJ...803...34B} (see their table 7). This conversion assumes a Salpeter IMF with mass range from $0.1-125~M_\odot$ and solar metallicity. We note that this is not the exact same IMF and metallicity used to calibrate $\eta_{\rm ion,a}$. However, we do not expect this to have a large impact on our results and indeed the discrepancy may be justified since the larger observed galaxies are likely to have somewhat higher metallicity than the smaller more abundant galaxies in atomic cooling haloes driving reionisation.

\begin{figure}
\includegraphics[width=88 mm]{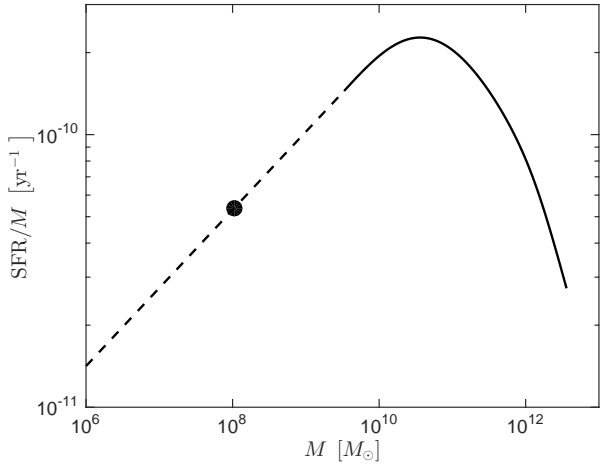}
\caption{\label{SFR_cal} SFR per halo mass as a function of halo mass computed with abundance matching and the UV LF at $z\approx5.9$ assuming a duty cycle of 10 per cent. The solid curve shows the masses corresponding to the observed data and the dashed line is a power-law extrapolation. The black circle corresponds to $M_{\rm a}$ at $z=6$. Note that we do not use this extrapolation below $M_{\rm a}$.   }
\end{figure}

\begin{figure}
\includegraphics[width=88mm]{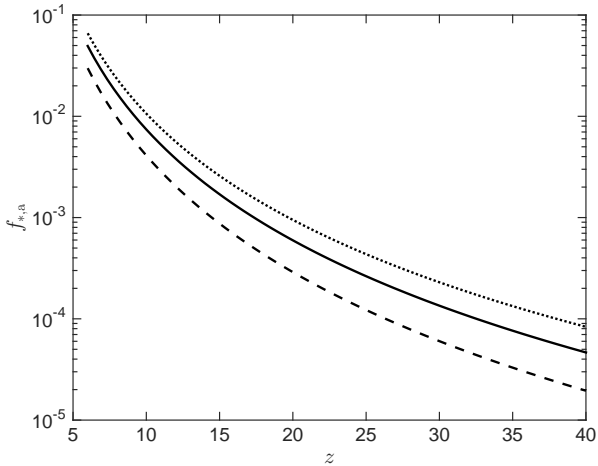}
\caption{ \label{f_star_vs_z} Star formation efficiency in atomic cooling haloes as a function of redshift in our redshift-dependent case computed with Eq. \ref{f_star_eqn}. The solid, dashed, and dotted curves are for a duty cycle, $\epsilon_{\rm duty}$, of 0.1, 0.5, and 0.02, respectively. }
\end{figure}

\begin{figure*}
\includegraphics[width=88mm]{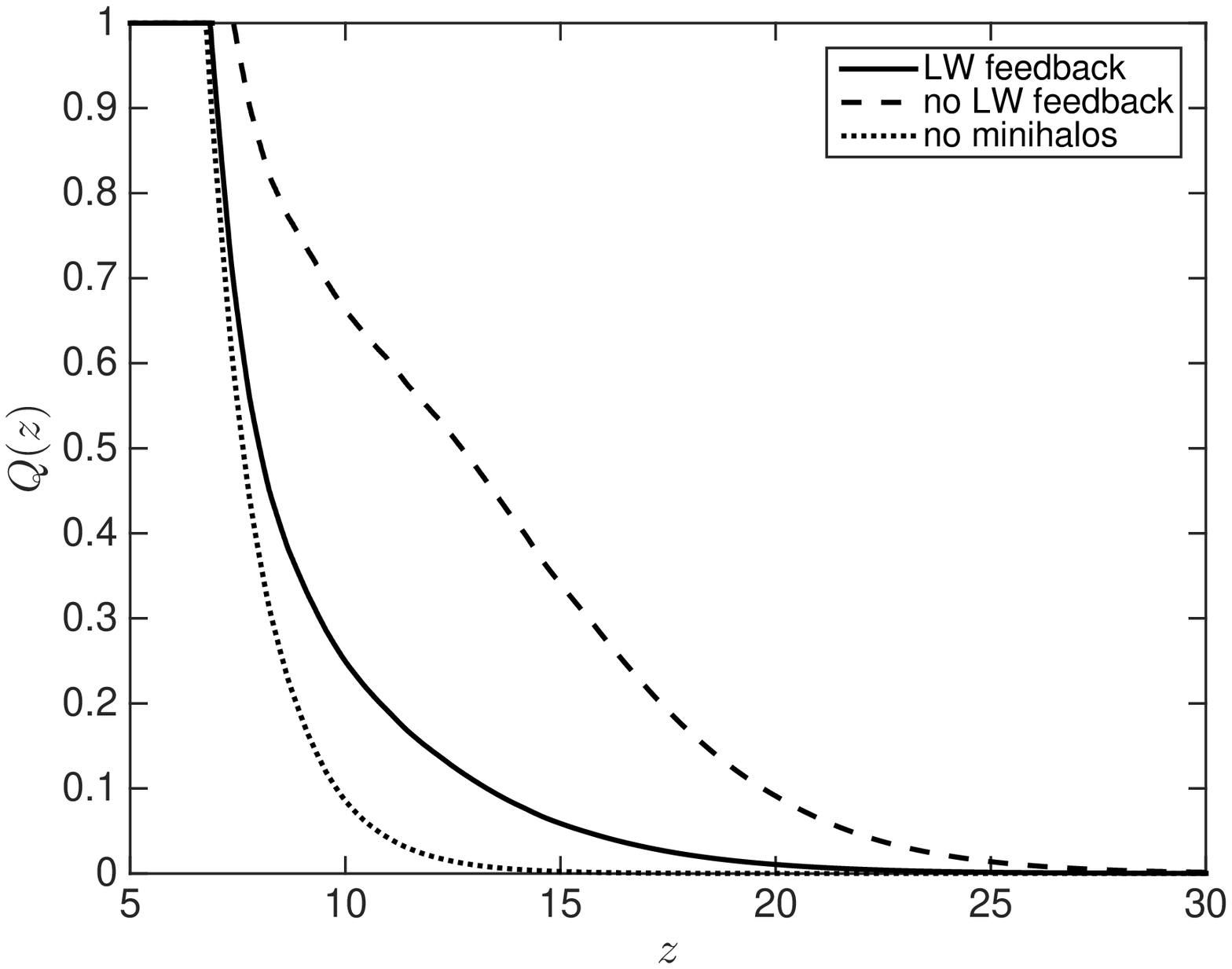}
\includegraphics[width=88mm]{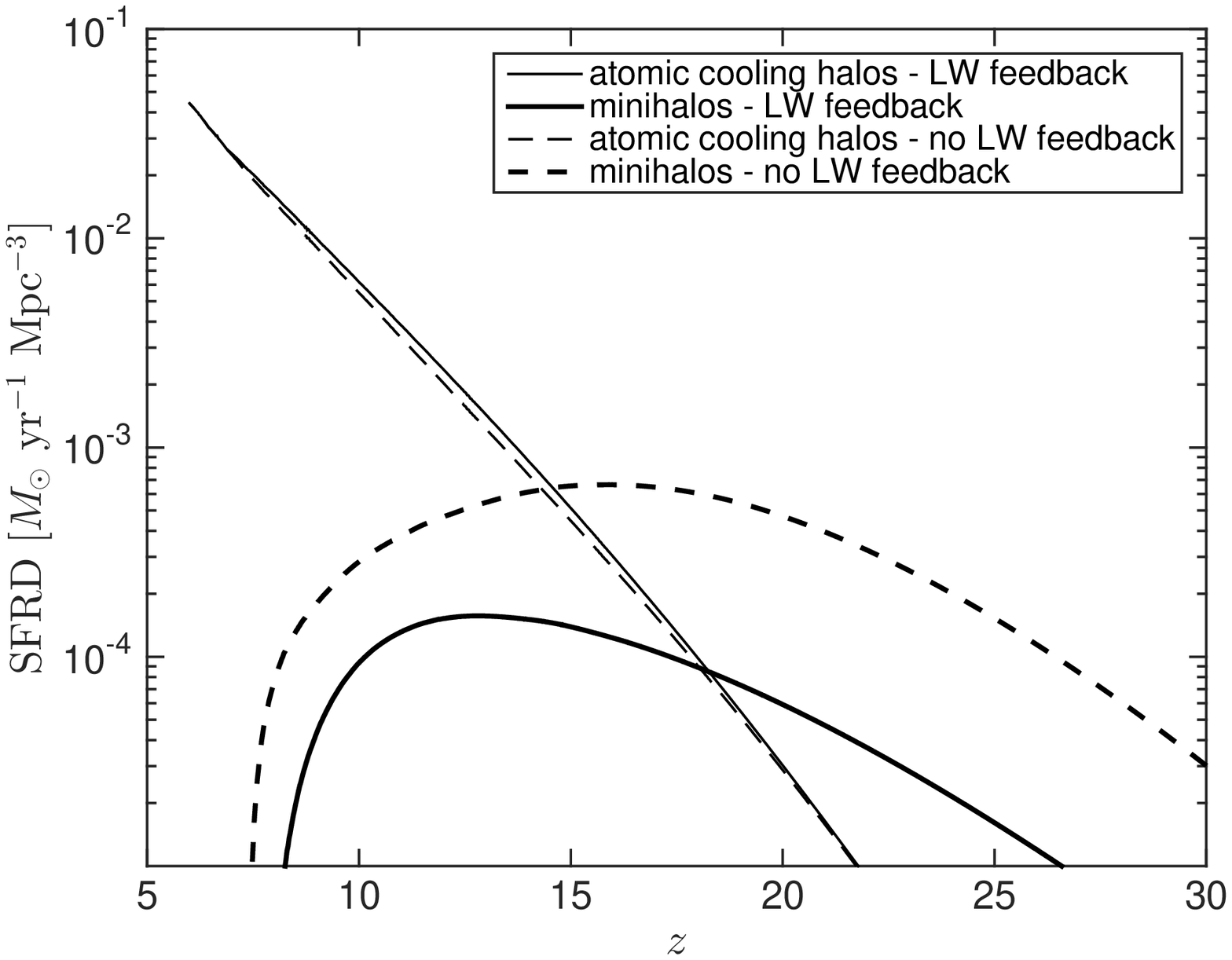}
\includegraphics[width=88mm]{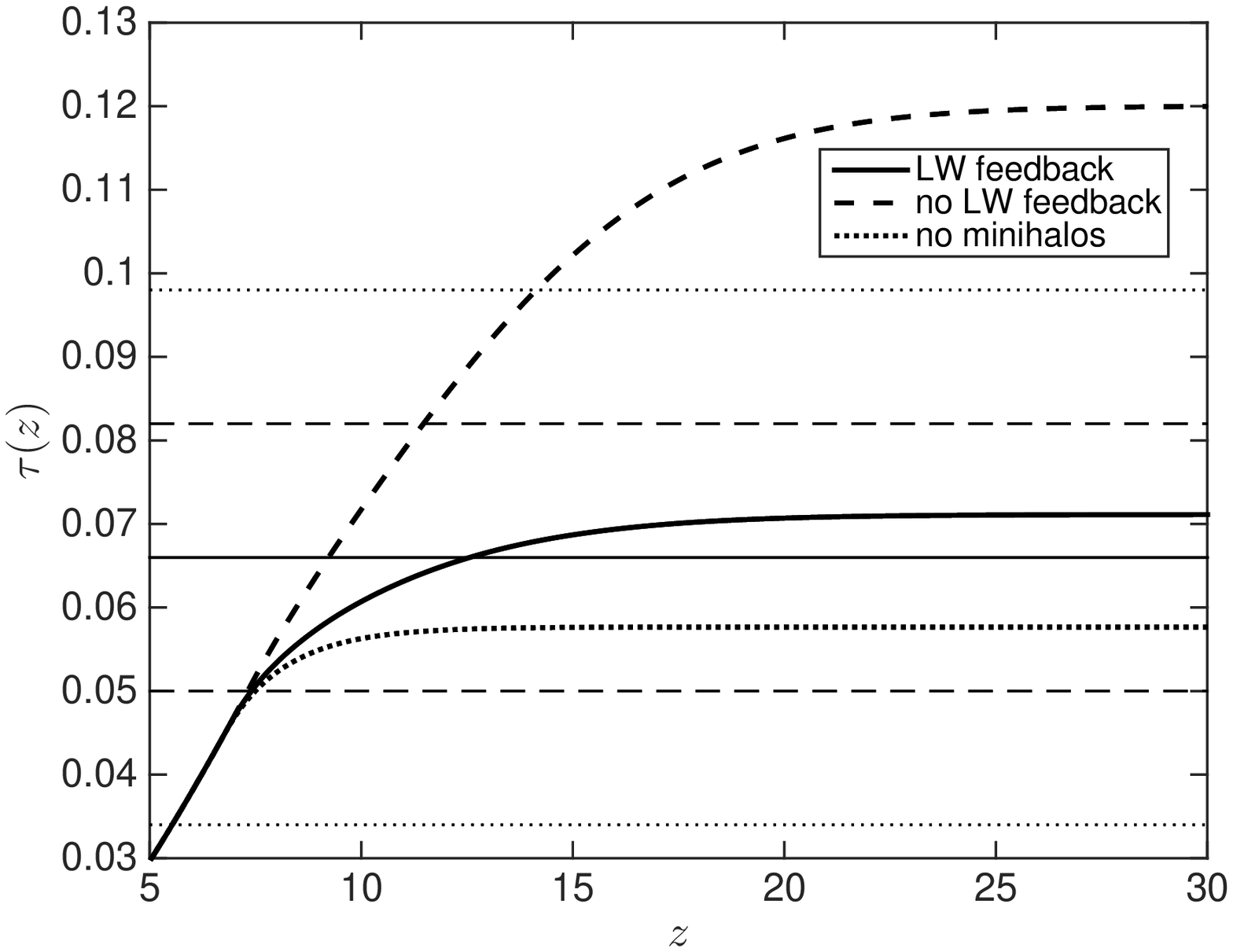}
\includegraphics[width=88mm]{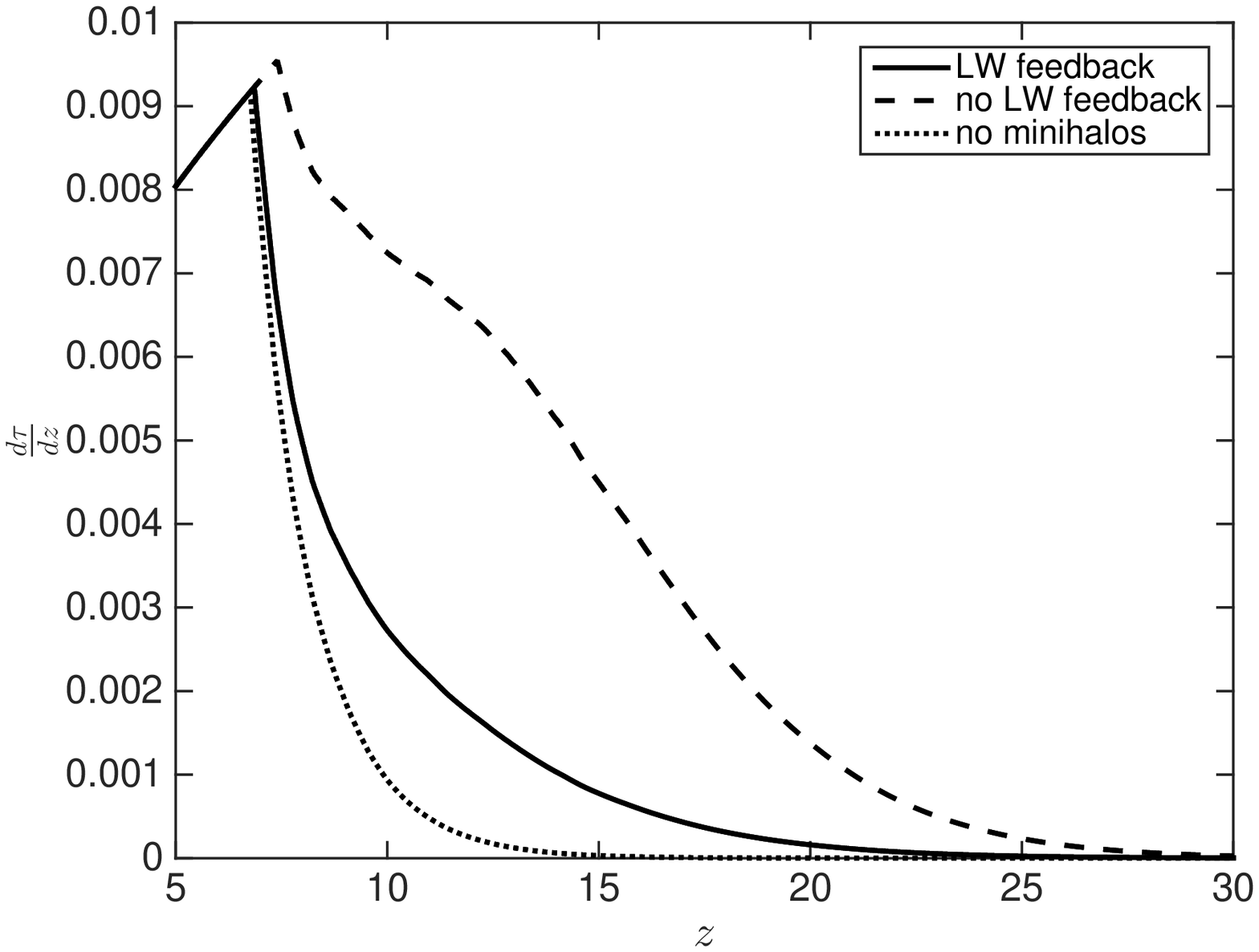}
\caption{ \label{fid_plots} Results of our fiducial reionisation model with the redshift-dependent $f_{\rm *, a}(z)$ shown in Figure \ref{f_star_vs_z}. We consider three different models for minihaloes:  $f_{\rm *, m}=0.001$ with (solid curves) and without  LW feedback (dashed curves), and a case without any contribution from minihaloes (dotted curves). We plot $Q(z)$, SFRD$(z)$ (for minihaloes and atomic cooling haloes), $\tau (z)$, and  $\frac{d\tau}{dz}$ for each of these models. In the $\tau(z)$ (lower left) panel, the thin solid, dashed, and dotted lines show the \emph{Planck} measurement, with the 1$\sigma$, and 2$\sigma$ errors, respectively. For the case without LW feedback, early ionisation leads to a high $\tau$ which is inconsistent with \emph{Planck}.}
\end{figure*}

In Figure \ref{SFR_cal}, we plot the SFR per halo mass as a function of halo mass. The abundance matching calculation indicates that star formation is most efficient in $\sim  {\rm a~ few}\times 10^{10}~M_\odot$ haloes and falls off at both lower and higher masses. The observed data extends down to a few times $10^9 M_\odot$. Below this value we use a power law extrapolation. We emphasise that this approach is conservative, because it diminishes global star-formation in atomic cooling haloes significantly towards higher redshifts, and leaves more room for ionising radiation from minihaloes. We use this SFR calibration to estimate the star formation efficiency in haloes above the atomic cooling threshold, $f_{\rm *, a}$, by taking the total instantaneous star formation rate divided by the total rate at which gas is being incorporated into virialised dark matter haloes
\begin{equation}
\label{f_star_eqn}
f_{\rm *,a} = \frac{\int_{M_{\rm a}}^{M_{\rm i}} dM {\rm SFR}(M)  \epsilon_{\rm duty} \frac{dn}{dM}}{ \rho_{\rm b} \frac{dF_{\rm coll, a}}{dt} }.
\end{equation}
Here SFR$(M)$ is the star formation rate in haloes of mass $M$ given by our abundance matching calculation. We note that the star formation efficiency does not change substantially if we substitute the limits of integration to correspond to haloes above the ionised IGM mass threshold. 
For the results presented below, we consider two different fiducial models of $f_{\rm *, a}$. In the first, redshift-dependent model, we assume that the SFR$(M)$ relationship is fixed with respect to redshift and compute how $f_{\rm *, a}(z)$ varies as a function of cosmic time from Eq. \ref{f_star_eqn} (see Figure \ref{f_star_vs_z}). We note that the reduction in $f_{\rm *,a}$ at high redshift is mostly due to the evolution of the halo mass function. In particular, the quantity $\left ( \int_{M_{\rm a}}^{M_{\rm i}} dM M \frac{dn}{dM} \right ) / \left ( \frac{dF_{\rm coll, a}}{dt} \right )$ has essentially the same redshift dependence as $f_{\rm *,a}$ in Figure \ref{f_star_vs_z}. At high redshifts, the mass in haloes relative to the rate of halo collapse is much lower. The exact ${\rm SFR}(M)$ relation shown in Figure \ref{SFR_cal}, sets the normalization of $f_{\rm *,a}(z)$, but only has a small effect on its redshift dependence. Figure \ref{f_star_vs_z} also shows the dependence of $f_{\rm *,a}$ on $\epsilon_{\rm duty}$. Changing the duty cycle from our fiducial value of 0.1 raises or lowers the normalization of the calibrated star formation efficiency and has a relatively small effect on the redshift dependence. 
In the redshift-independent model we assume $f_{\rm *, a}=0.05$, which is the value we compute with Eq.   \ref{f_star_eqn} at $z=6$.

 \subsection{Metal enrichment}
Up to this point, we have made the simplifying assumption that minihaloes host only Pop III star formation and atomic cooling haloes host only Pop II star formation. However, in reality, supernovae winds enrich some minihaloes with metals enabling Pop II star formation. Conversely, strong LW feedback can lead to the formation of atomic cooling haloes without metals, resulting in Pop III star formation. To estimate the impact of these effects, we consider a simple idealized model of metal enrichment where the smallest star-forming haloes are enriched by Pop III supernovae and subsequently form larger enriched haloes. Specifically, we assume that Pop III star formation occurs with efficiency $f_{\rm *,III}$ only between $M_{\rm m}$ and $2M_{\rm m}$, given by Eq. \ref{mini_mass}. Pop II star formation is assumed to form in all larger haloes with efficiency $f_{\rm *,II}$. In haloes forming Pop II stars we adopt the same parameters ($f_{\rm esc}$, $\eta_{\rm ion}$, $\eta_{\rm LW}$, etc.) as we did for atomic cooling haloes described above (and utilize both the redshift-independent and redshift-dependent star formation efficiency models). Similarly, for Pop III star-forming haloes we adopt the parameters for minihaloes described above. Operationally, our model is computed by changing the upper limit of integration in Eq. \ref{minifrac} and the lower limit of integration in Eq. \ref{atom_frac} to 2$M_{\rm m}$. When LW feedback is included, if the LW background is strong enough to suppress star formation in all minihalos (i.e. if the value of $J_{\rm LW}$ causes $M_{\rm m}$ in Eq. \ref{mini_mass} to be greater than $M_{\rm a}$), we assume that Pop III stars form in atomic cooling haloes between $M_{\rm a}$ and 2$M_{\rm a}$ (with efficiency $f_{\rm *,III}$) and Pop II stars form in larger haloes. We model this by changing the limits of integration in Eq. \ref{minifrac} to $M_{\rm a}$ and $2M_{\rm a}$, and the lower limit of integration in Eq. \ref{atom_frac} to $2M_{\rm a}$. 
Note that for simplicity we have chosen a factor of two in the mass range of halos hosting Pop III stars. This value is not expected to be a precise description of the true mass range. However, a factor of two is not unreasonable since two halos with masses equal to the minimum star forming mass could merge to produce a metal enriched halo. 

Note that this treatment is not self-consistent for very low values of $f_{\rm *, III}$ because there will not be enough metals produced in haloes smaller than $2M_{\rm m}$ to lead to a transition to Pop II stars. 
Thus, for low $f_{\rm *,III}$, our model is conservative in the sense that we shut-off Pop III star formation 
in smaller haloes earlier than expected, leading to weaker constraints on the Pop III star formation efficiency. 
For reference, $f_{\rm *,III}\approx10^{-3}$ in a $10^6~M_\odot$ minihalo corresponds to $\sim 150~M_\odot$ of stars, which could be enough to produce a pair instability supernova and enrich gas to the levels required for Pop II star formation. 

We also point out that our simple model focuses on ``self-enrichment'' of haloes by their progenitor haloes and neglects the impact of minihalo enrichment from winds emitted by nearby larger haloes. This works in the opposite direction of the inconsistency described above, enriching some small haloes that are metal-free in our model. Despite these shortcomings, we expect our simple treatment to give a rough indication of the impact of metal enrichment on the reionisation model described above.

\section{Results}
We present the results of our semi-analytic reionisation model in the following three subsections. First, we present results from our model with the simplifying assumption that minihaloes contain only Pop III stars and atomic cooling haloes contain only Pop II stars. In Section 3.2, we show how these results could change due to metal enrichment in minihaloes with the prescription described in Section 2.4. Finally, in Section 3.3 we present limits on the total density of Pop III stars formed across cosmic time with and without LW feedback and minihalo metal enrichment.

\subsection{No minihalo metal enrichment}
\begin{figure}
\includegraphics[width=88mm]{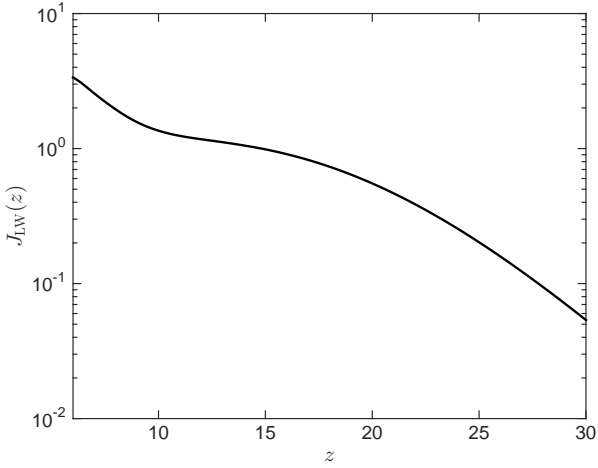}
\caption{ \label{J_plot} The LW background for our fiducial redshift-dependent $f_{\rm *,a}$ model with $f_{\rm *,m}=0.001$ in units of $10^{-21}~{\rm ergs~s^{-1}~cm^{-2}~Hz^{-1}~Sr^{-1}}$. }
\end{figure}

In Figure \ref{fid_plots}, we show results for our reionisation model with redshift-dependent $f_{\rm *, a}$ and the fiducial parameters discussed above. We plot $Q(z)$, SFRD$_{\rm a}(z)$, SFRD$_{\rm m}(z)$,  the CMB optical depth, $\tau(z)$, and $\frac{d\tau}{dz}$, along with the observational limits from \emph{Planck}. We consider three cases: $f_{\rm *, m}=0.001$ without LW feedback, $f_{\rm *, m}=0.001$ including LW feedback, and $f_{\rm *, m}=0$. For the case with LW feedback, we plot $J_{\rm LW}(z)$ in Figure \ref{J_plot}. The choice of star formation efficiency plotted represents moderate formation of massive Pop III stars (for $f_{\rm *, m}=0.001$, a typical $10^6~M_\odot$ minihalo would form $\sim150~M_{\odot}$ in stars).  We note that the total SFRD$_{\rm a}$ at $z=6$ is similar to that inferred from the observations of \cite{2015ApJ...803...34B}  (see their figure 18) since we calibrate with their UV LF and their assumed limiting magnitude roughly corresponds to our $M_{\rm i}$. However at higher redshifts our SFRD is higher somewhat higher because we include the contribution from fainter galaxies.

We compute the optical depth assuming helium is singly ionised at the same time as hydrogen and doubly ionised instantaneously at $z=3$. This leads to
\begin{equation}
\tau(z) = \int_0^z dz' \frac{c(1+z')^2}{H(z')}Q(z') \sigma_{\rm T} \langle n_{\rm H} \rangle \left (1+  \eta_{\rm He} \frac{Y}{4X} \right ), 
\end{equation}
where $\sigma_{\rm T}$ is the Thompson scattering cross section, $Y=0.24$ and $X=0.76$ are the helium and hydrogen mass fractions, and $\eta_{\rm  He}=1$ for $z'>3$ and $\eta_{\rm  He}=2$ for $z'\leq 3$.

It is clear from Figure \ref{fid_plots} that without LW feedback, a Pop III star formation efficiency of $f_{\rm *, m}=0.001$ in minihaloes is inconsistent with the \emph{Planck} optical depth. Partial reionisation begins early, leading to a $\tau$ more than $3\sigma$ higher than the \emph{Planck} measurement. LW feedback reduces the SFRD in minihaloes by roughly an order of magnitude below $z=30$. This leads to $\tau=0.071$, which is consistent with \emph{Planck} and leads us to the main conclusion of this paper: if massive Pop III stars form in minihaloes with moderate efficiency, LW feedback, metal enrichment, or a comparable other suppression of PopIII star-formation is necessary to prevent $\tau$ from being too high compared to \emph{Planck} data.

\begin{figure*}
\includegraphics[width=88mm]{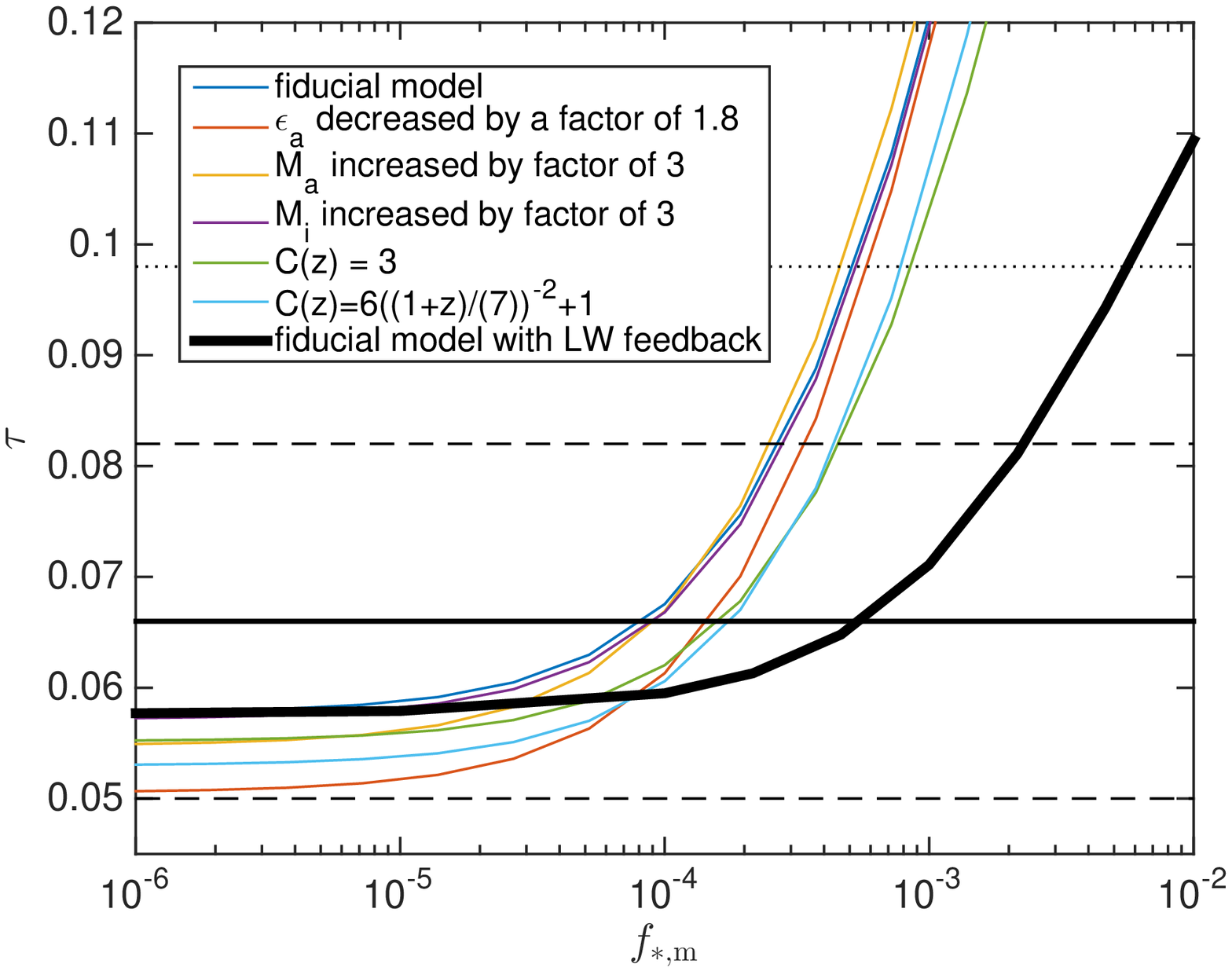}
\includegraphics[width=88mm]{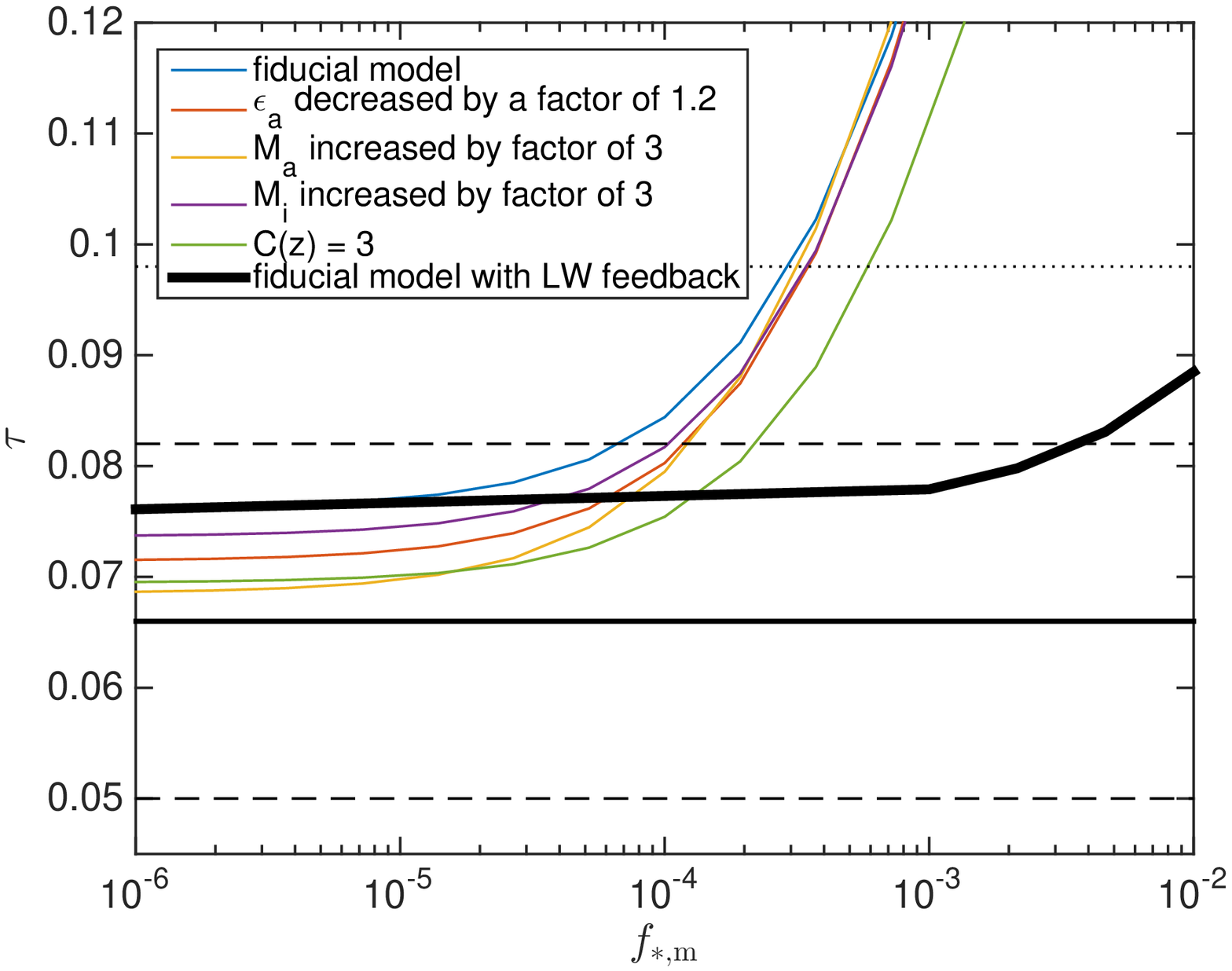}
\caption{\label{tau_vs_f}  The optical depth due to electron scattering vs. minihalo star-formation efficiency for two different parameterisations of our reionisation model, and for various different parameter values. For comparison, we plot the $\tau$ measured by \emph{Planck} (solid black line) and the 1$\sigma$ and 2$\sigma$ error bars (dashed and dotted black lines). For all cases we have assumed $f_{\rm esc, m}=0.5$ and $\eta_{\rm ion,m}=80000$. The redshift-dependent $f_{\rm *, a}$ model (see Figure \ref{f_star_vs_z}) is plotted in the left panel and the redshift-independent, $f_{\rm *, a}=0.05$, model is plotted in the right panel. For the cases without LW feedback all parameters have been varied in the direction which decreases $\tau$, providing a conservative upper limit on $f_{\rm *, m}$. The variations in $\epsilon_{\rm a}$ represent the largest possible decrease without causing reionisation to complete at $z<6$ (in the right panel, we do not include the factor of 3 change in $C(z=6)$ because this delays reionisation to $z<6$). }
\end{figure*}

\begin{figure*}
\includegraphics[width=88mm]{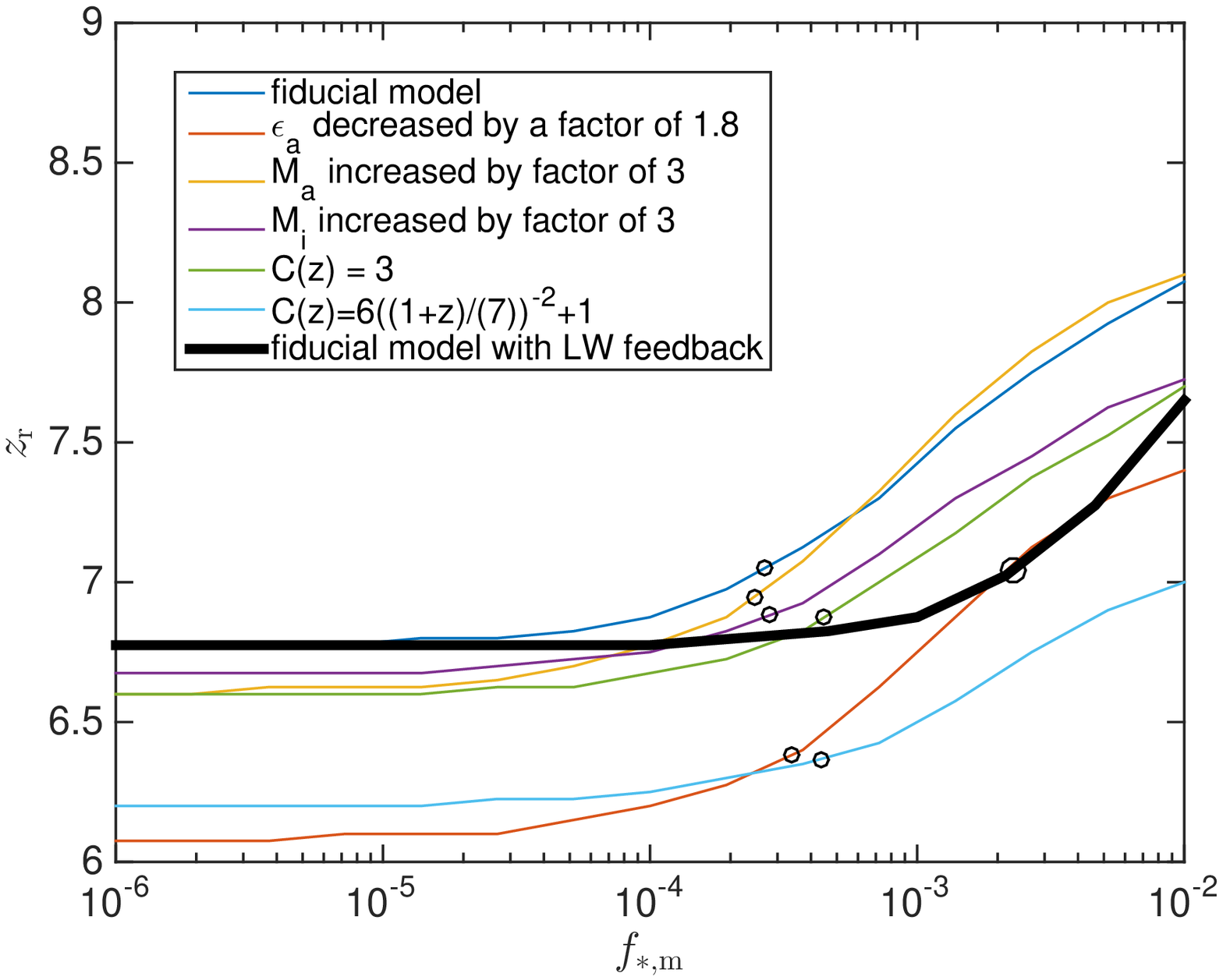}
\includegraphics[width=88mm]{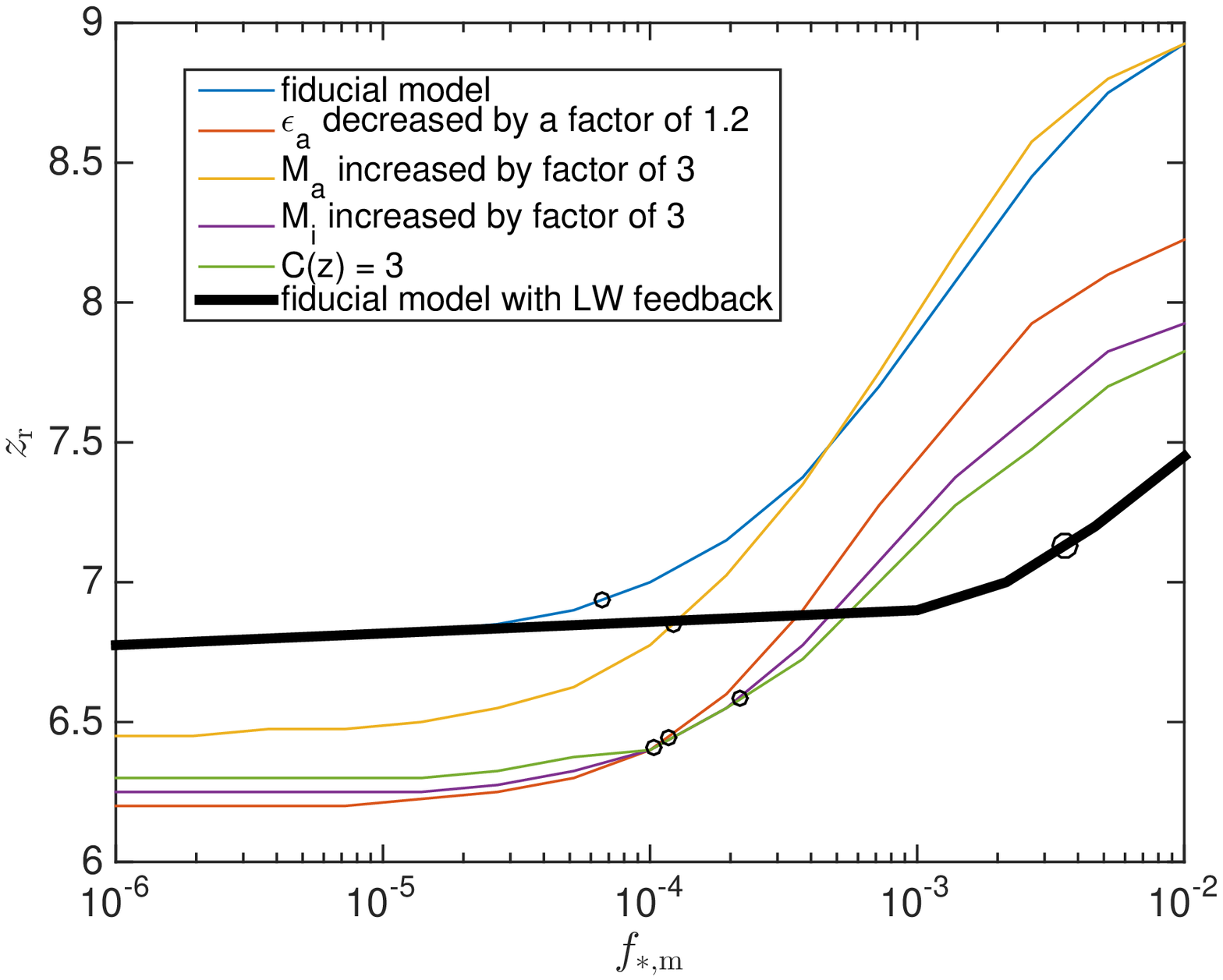}
\caption{\label{zr_vs_f}  Same as Figure \ref{tau_vs_f}, but showing the redshift when reionisation is completed, $z_{\rm r}$, instead of $\tau$. The black circles denote the star formation efficiency corresponding to the 1-$\sigma$ \emph{Planck} optical depth limits. }
\end{figure*}

We emphasise that our main conclusions are not sensitive to the exact choices of fiducial parameters.
To demonstrate this, we vary each of the parameters in our model associated with atomic cooling haloes, subject to the constraint that reionisation is essentially complete ($Q\sim 1$) by $z\approx6$ \citep{Mesinger2010}, and determine how efficiently minihaloes can produce Pop III stars and still be consistent with \emph{Planck}.  In Figure \ref{tau_vs_f}, we plot $\tau$ as a function of $f_{\rm *, m}$ (assuming $\eta_{\rm ion,m}=80000$ and $f_{\rm esc, m}=0.5$). We plot this both for our redshift-independent and redshift-dependent $f_{\rm *, a}$ models without LW feedback and individually vary the other physical parameters (besides those associated with minihaloes): $\epsilon_{\rm a}$, $M_{\rm a}$, $M_{\rm i}$, and $C(z)$. We vary these parameters in the direction which lowers $\tau$, allowing a higher $f_{\rm *, m}$. Thus, we present conservative upper limits on the efficiency of Pop III stars that can form in minihaloes without LW feedback. We also show how $\tau$ depends on $f_{\rm *, m}$ in the fiducial case with LW feedback. Additionally, we plot the redshift when reionisation is complete, $z_{\rm r}$, for the same models in Figure \ref{zr_vs_f}. Note that the values plotted for a lower $\epsilon_{\rm a}$ are the lowest possible, since lowering them any more than this prevents reionisation from completing before $z=6$.

Overall we find broadly consistent results for the two different parameterisations of our reionisation model. The redshift-dependent $f_{\rm *, a}$ model with fiducial parameters and no LW feedback requires $f_{\rm *, m} \lesssim 3\times 10^{-4}$ to be consistent to within 1$\sigma$ of the \emph{Planck} $\tau$ . This corresponds to $\lesssim 50 M_\odot$ of stars in a typical $10^6M_\odot$ dark matter halo. Thus, more massive Pop III stars forming in most minihaloes without LW feedback is inconsistent with \emph{Planck}. We find that changing the fiducial parameters does not have a large effect on these results. Even for the variations in $C(z)$, which we regard as extremely conservative, the limits on $f_{\rm *,m}$ only change by about a factor of 2.
When LW feedback is included, $\tau$ is consistent with \emph{Planck} for $f_{\rm *, m} \lesssim 0.002$. This would permit massive Pop III stars in typical minihaloes, before LW feedback prevents star formation.

We find similar, but even more severe constraints on $f_{\rm *,m}$ in the redshift-independent $f_{\rm *, a}$ model without LW feedback. However, when LW feedback is included, because the background is higher at early times due to the greater $f_{\rm *, a}$, the star formation efficiency in minihaloes can be $f_{\rm *, m}\lesssim0.003$ without violating the \emph{Planck} constraints. Note that we do not show the case with  $C(z)= 6 \left ( \frac{1+z}{7}  \right )^{-2} + 1$ because this leads to a reionisation history where reionisation is not complete until after $z=6$.

\subsection{Impact of minihalo metal enrichment}
\begin{figure*}
\includegraphics[width=88mm]{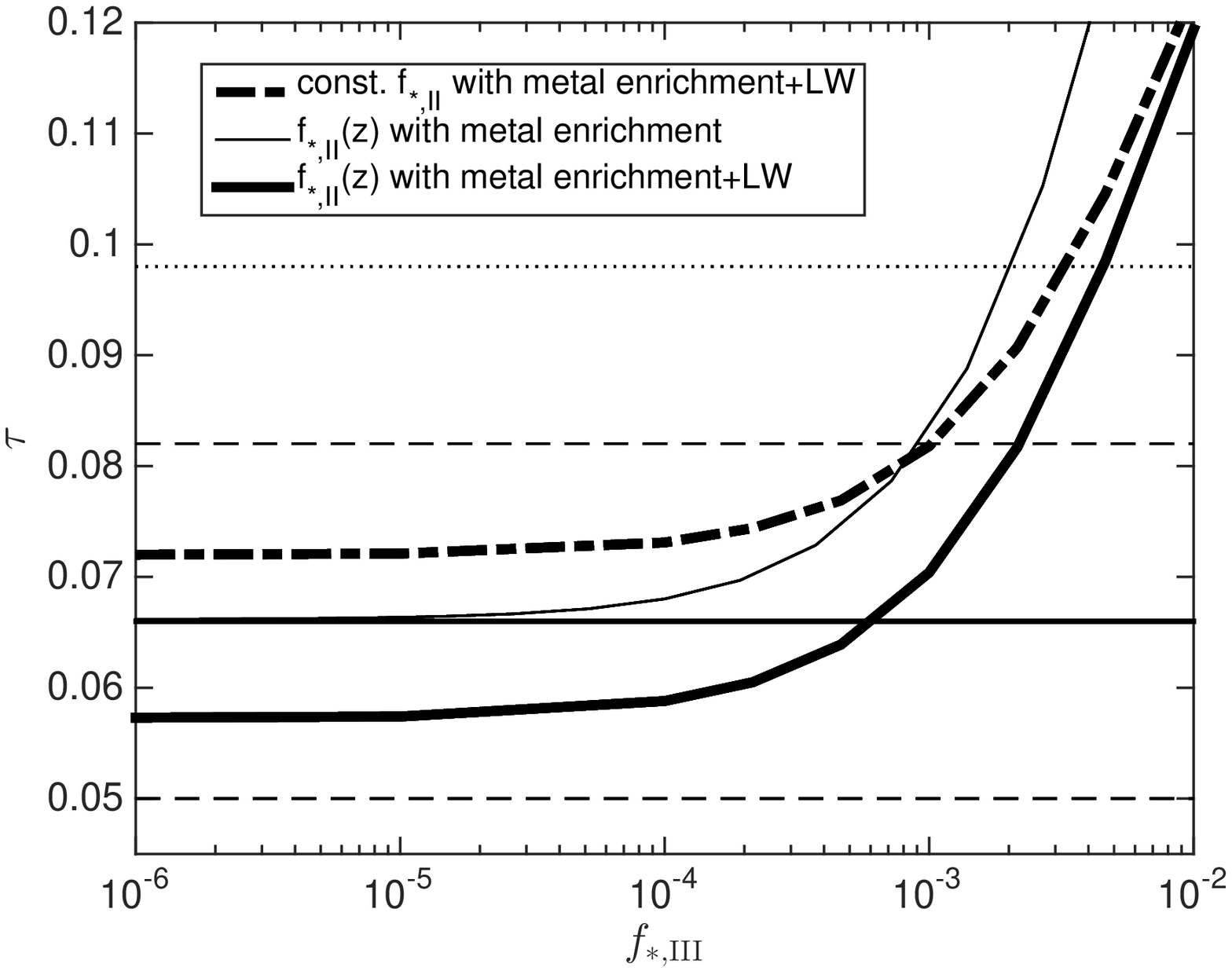}
\includegraphics[width=88mm]{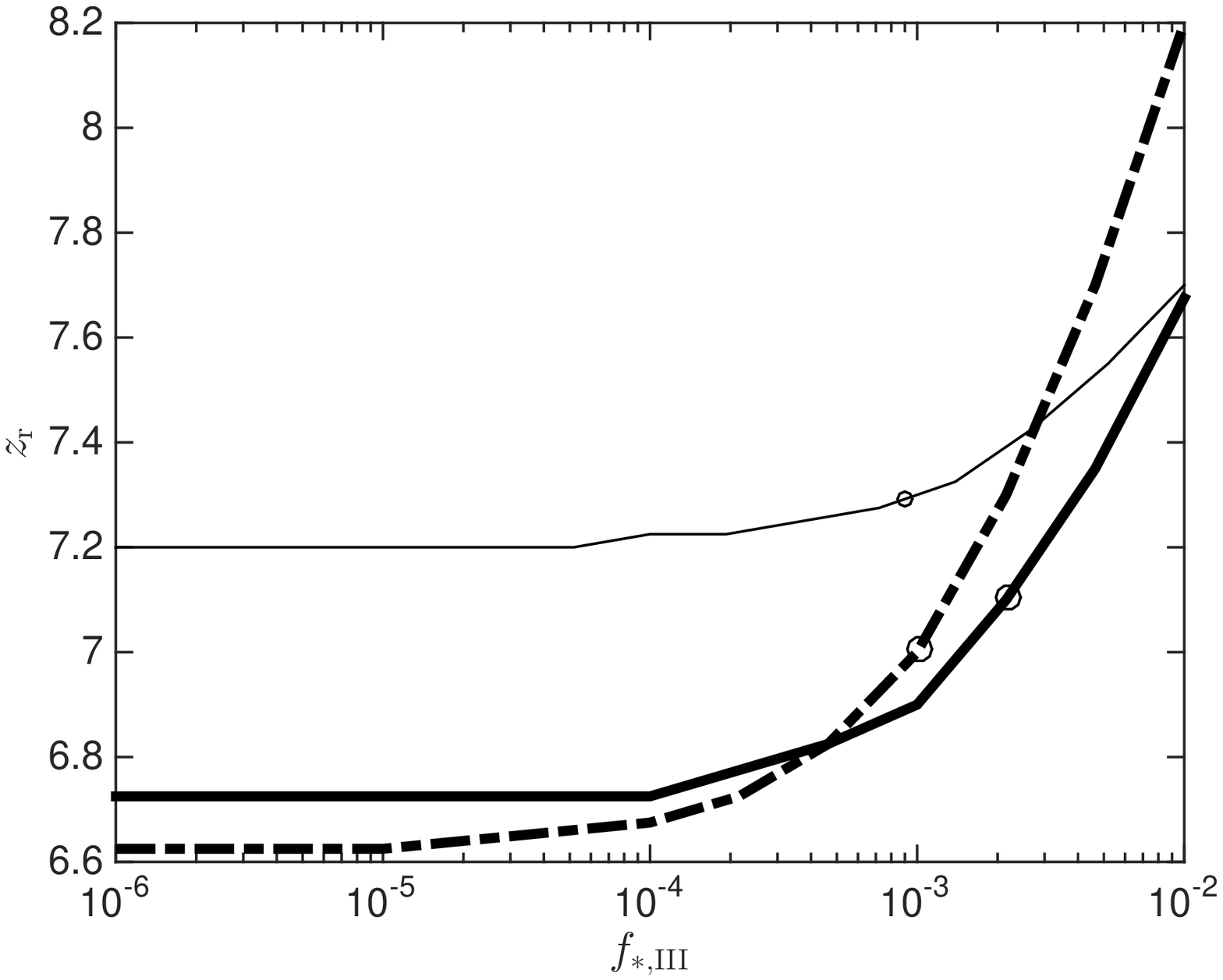}
\caption{\label{metal_tau_vs_ep} The CMB optical depth and redshift of reionisation for our semi-analytic model with metal enrichment. The horizontal solid, dashed, and dotted lines show the \emph{Planck} $\tau$ measurement, 1-$\sigma$, and 2-$\sigma$ limits, respectively. The black circles in the right panel indicate the star formation efficiency corresponding to the 1-$\sigma$ \emph{Planck} limits. We do not include the redshift-independent $f_{\rm *,II}$ case without LW feedback, because unphysical Pop II star formation occurs at very high redshift in minihaloes leading to an artificially high optical depth.  }
\end{figure*}

In Figure \ref{metal_tau_vs_ep}, we plot $\tau$ and $z_{\rm r}$ as a function of Pop III star formation efficiency, $f_{\rm *,III}$, for our reionisation model including the treatment of metal enrichment described in Section 2.4 with and without LW feedback. We find that for both the redshift-dependent and redshift-independent $f_{\rm *,II}$ models, when LW feedback and metal enrichment are included, the \emph{Planck} 1-$\sigma$ limit on $\tau$ corresponds to a Pop III star formation efficiency of $f_{\rm *,III}\approx10^{-3}$. For the redshift-independent $f_{\rm *, II}$ case with LW feedback and metal enrichment, the limit on Pop III star formation efficiency goes down by a factor of $\sim 4$ compared to the case with LW feedback and no metal enrichment. This is because our treatment of metal enrichment causes Pop III stars to form in the smallest atomic cooling haloes, increasing the total amount of ionizing photons produced for sufficiently high Pop III star formation efficiency. 

For the redshift-dependent $f_{\rm *,II}$ case with LW feedback, we find that the limits on Pop III star formation efficiency are similar with or without metal enrichment. With $f_{\rm *,III}\sim10^{-3}$, we find that in both cases, for $z \lesssim 20$, the LW feedback sets the minimum minihalo mass to be roughly half the atomic cooling mass, which leads to Pop III star formation in the same mass range with or without metal enrichment and explains the close similarity in $\tau$. In general, we find that the effects of metal enrichment plus LW feedback do not greatly change the constraints on Pop III star formation efficiency obtained with LW feedback alone.

We note that in reality metal enrichment will not lead to a global change in the mass range of halos that host Pop III stars. Instead there will be a complex interplay between radiative feedback and metal enrichment leading to the mass range varying strongly as a function of position. As such, we caution the reader that the results in this subsection are meant only to give a rough indication of the possible impact of metal enrichment on the constraints on Pop III stars.

\subsection{Limits on total density of Pop III stars}
Next, we examine how the \emph{Planck} optical depth limits the total density of Pop III stars formed over cosmic time. We compute this by integrating $SFRD_{\rm m}(z)$ ($SFRD_{\rm III}(z)$ in the case with metal enrichment) with respect to cosmic time. 

\begin{figure*}
\includegraphics[width=88mm]{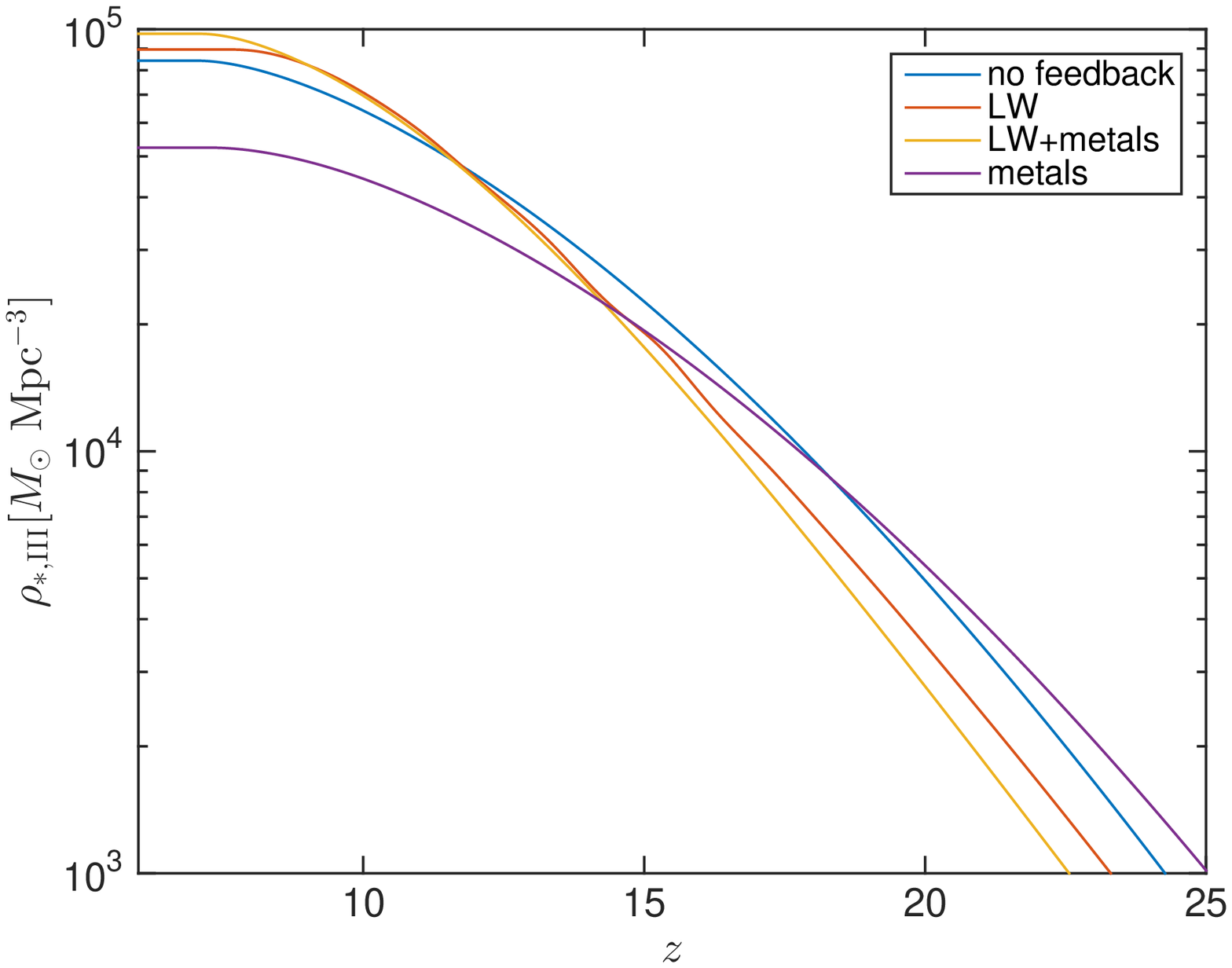}
\includegraphics[width=88mm]{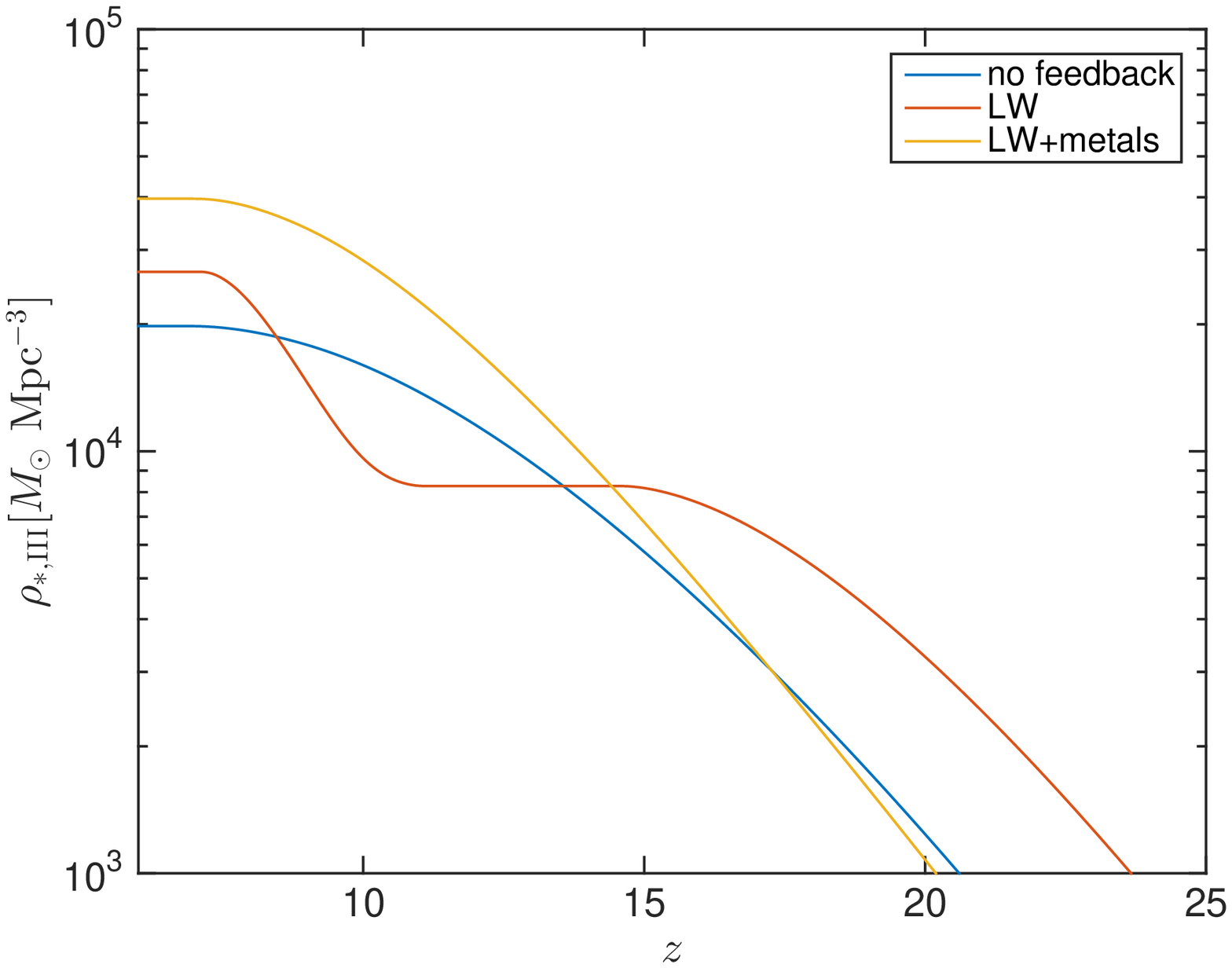}
\caption{\label{rho_popIII}   The total cumulative density of Pop III stars as a function of redshift for Pop III star formation efficiency corresponding to the 1-$\sigma$ \emph{Planck} upper limits, including LW feedback and/or metal enrichment. The left (right) panel shows the results for the redshift-dependent (redshift-independent) $f_{\rm *,a}$ model.  }
\end{figure*}

In Figure \ref{rho_popIII}, we plot the cumulative density of Pop III stars, $\rho_{\rm *, III}$, with $f_{\rm *, m}$ ($f_{\rm *, III}$ in cases with metal enrichment) corresponding to the 1-$\sigma$ limits on the \emph{Planck} optical depth. We show results for the redshift-independent and redshift-dependent $f_{\rm *,a}$ models with and without LW feedback and our treatment of metal enrichment. Interestingly, we find that including LW feedback and/or metal enrichment does not have a large impact on the limits of the total number of Pop III stars produced. For the redshift-dependent and redshift-independent $f_{\rm *,a}$ cases, we find total limiting Pop III densities of $\sim 10^5~M_\odot~{\rm Mpc^{-3}}$ and $\sim 3 \times 10^4~M_\odot~{\rm Mpc^{-3}}$, respectively. Including or not including LW feedback and metal enrichment does not change this result by more than a factor of a few. LW feedback or prompt metal-enrichment reduce the number of mini-halos that can form Pop III stars, allowing the efficiency of Pop III star formation to rise in those halos, but ultimately the strict new limits on $\tau$ only permit a small number of Pop III stars to form overall. 

{\bf We note that in our models with LW feedback, but no metal enrichment, Pop III star formation can be completely suppressed when the LW background is sufficiency high. This can be seen in the right panel of Figure \ref{rho_popIII}, where Pop III star formation is halted at $z\approx15$ (however it resumes at $z\approx10$ when the LW background decreases). For the other models, Pop III star formation is only completely stopped by the completion of reionization.}

\section{Limits on Pop III BH seed growth}
If massive Pop III stars are produced in minihaloes, some will end their lives as black holes. As these black holes grow, they are expected to produce X-rays which provide an additional source of ionising radiation not included in the model described above. In this section, we perform a simple calculation to estimate the impact of black hole accretion on the IGM.

To estimate the total mass of black holes produced, we use the fiducial redshift-dependent $f_{\rm *,a}$ model described above, including LW feedback and $f_{\rm *, m}=0.001$. We assume that 10 per cent of the stellar mass formed in minihaloes ends up as black holes (computed with the SFRD in Eq. \ref{mini_sfrd}). This is approximately the fraction obtained for a Salpeter IMF with mass limits of 1 $M_\odot$ and 100 $M_\odot$ assuming that stars with with initial mass between 40  $M_\odot$ and 100  $M_\odot$ collapse directly to black holes. Note that a more top heavy IMF could produce a somewhat larger fraction. 

We assume that black holes grow at the Eddington limit a fraction $\epsilon_{\rm BH, duty}$ of the time with radiative efficiency $\epsilon_{\rm r}=0.1$. For the entire population of black holes this results in an accretion rate  
\begin{equation}
\frac{d\rho_{\rm BH}}{dt} = 2.2 \times 10^{-9} \epsilon_{\rm BH, duty} \frac{1-\epsilon_{\rm r}}{\epsilon_{\rm r}} \rho_{\rm BH} ~ {\rm yr}^{-1},
\end{equation}
 where $\rho_{\rm BH}$ is the total comoving black hole density. If the black hole density reaches $\rho_{\rm BH}=10^5 ~M_\odot {\rm Mpc^{-3}}$, we turn off accretion by hand (mimicking inefficient growth or a self-regulation; \citealt{2012MNRAS.425.2974T}) to prevent the density from greatly exceeding that of SBMHs in the local universe.

We compute the effect of the black hole growth on the IGM by taking the number of ionisations per volume per time as
\begin{equation}
\frac{dn_{\rm ion}}{dt} = \frac{f_{\rm ion}\epsilon_{\rm r} c^2}{E_\gamma} \frac{d\rho_{\rm BH}}{dt}  ,
\end{equation}
where $E_\gamma=13.6$ eV is energy required to ionise a hydrogen atom. An X-ray produced through black hole accretion will ionise a hydrogen or helium atom producing a high energy electron. A fraction of this electron's energy, $f_{\rm ion}$, will go into producing additional ionisations. The value of $f_{\rm ion}$ depends on the energy of the electron and the ionised fraction of the IGM. We estimate $f_{\rm ion}$ by interpolating the results of \cite{2010MNRAS.404.1869F} and assuming a typical electron energy of 1 keV. For an ionised fraction close to zero this gives us $f_{\rm ion} \approx 0.4$ and is reduced to nearly zero as the ionised fraction approaches unity. Using this rate of ionising photon production and the fiducial $C(z)$ described above, we compute the ionised fraction from black hole accretion alone, $Q_{\rm X}(z)$. We assume that because of the large mean free path of X-rays, the IGM is uniformly ionised, as opposed to having an ionised bubble topology \citep[e.g.][]{Oh01}. Since we are only performing a rough estimation of the impact of black hole accretion, we do not attempt to self-consistently model the combination of X-rays and UV photons from stars simultaneously. Note that we also do not model possible self-regulation of black hole growth via X-ray feedback, as was explored in \cite{2012MNRAS.425.2974T}.

\begin{figure*}
\includegraphics[width=88mm]{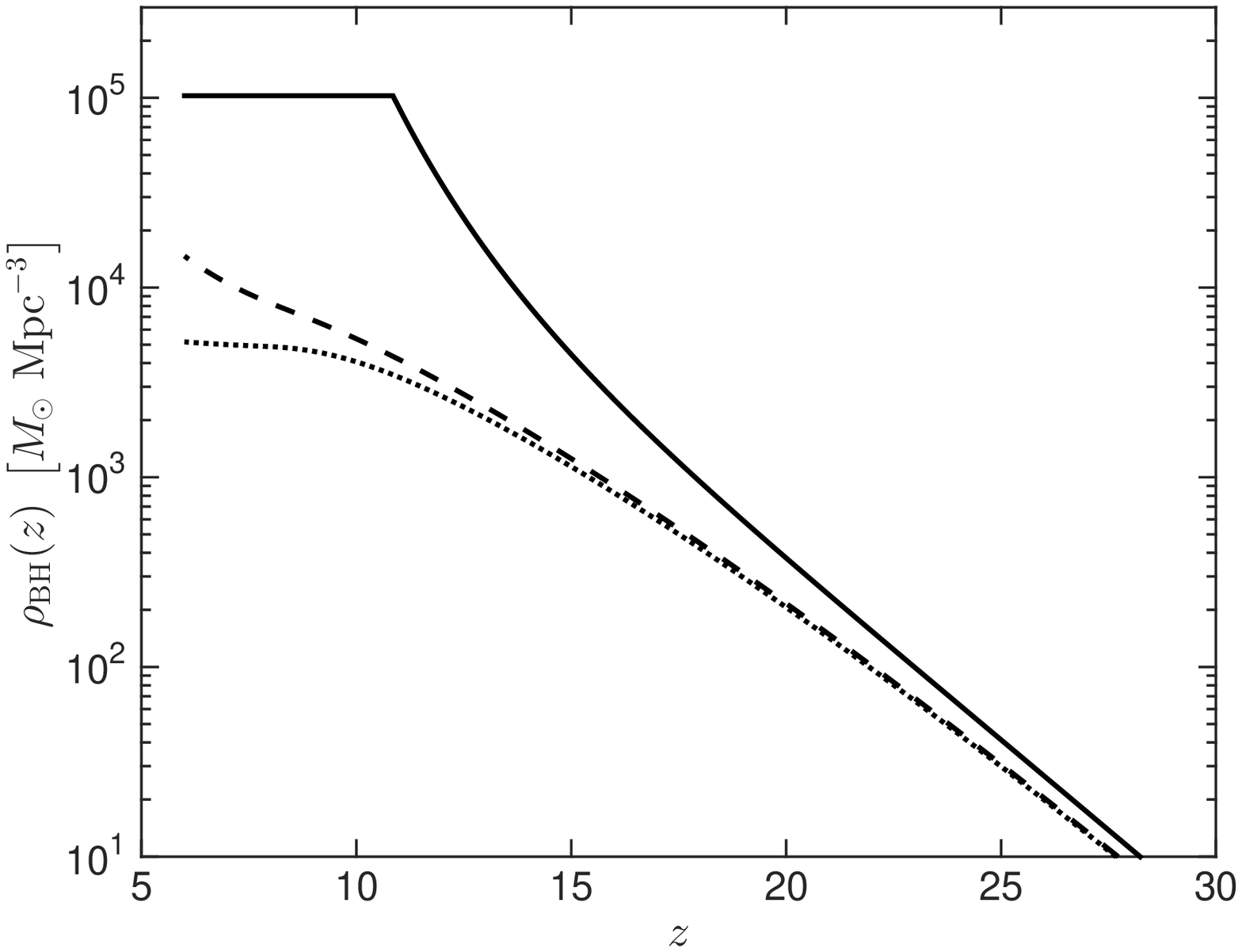}
\includegraphics[width=88mm]{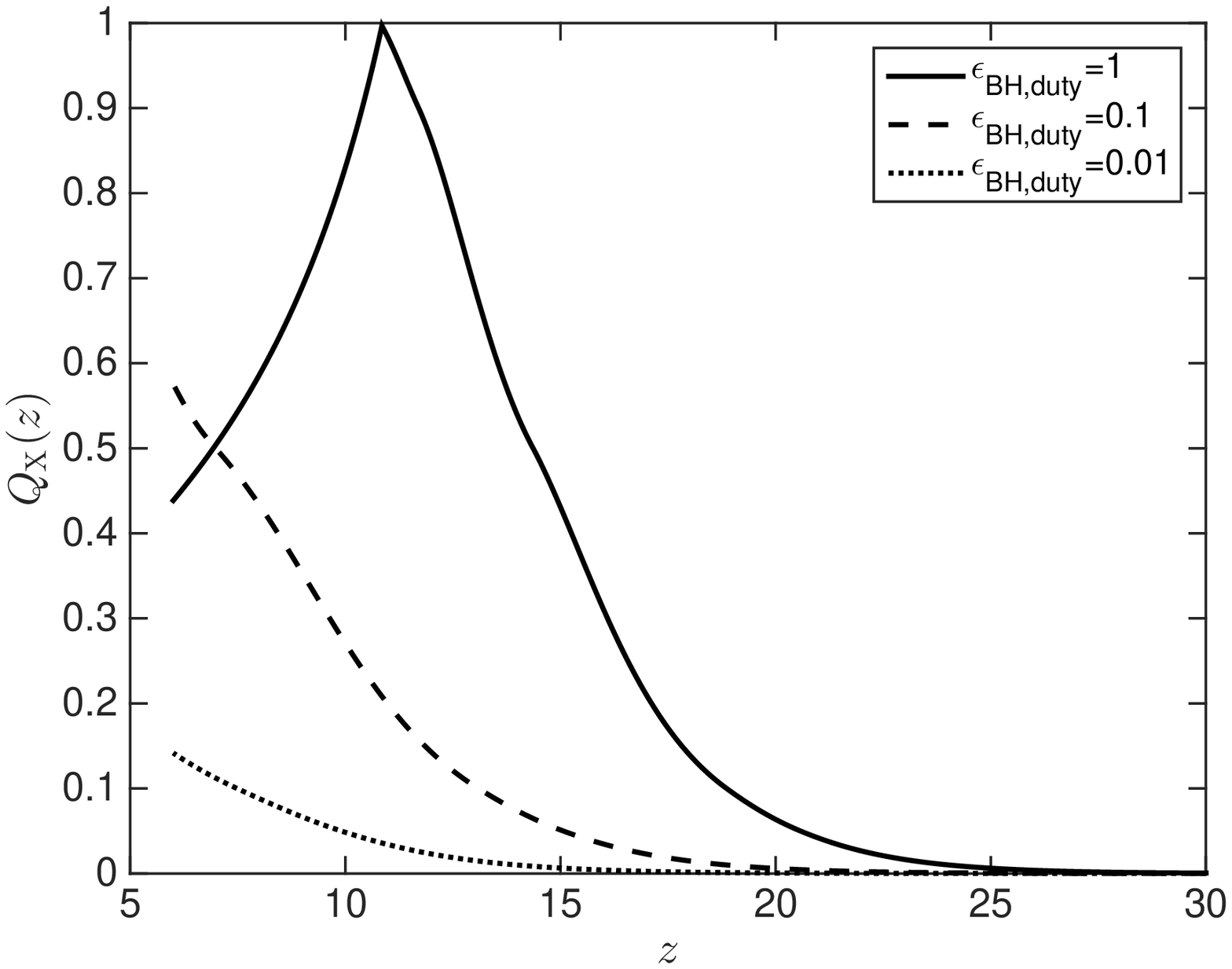}
\includegraphics[width=88mm]{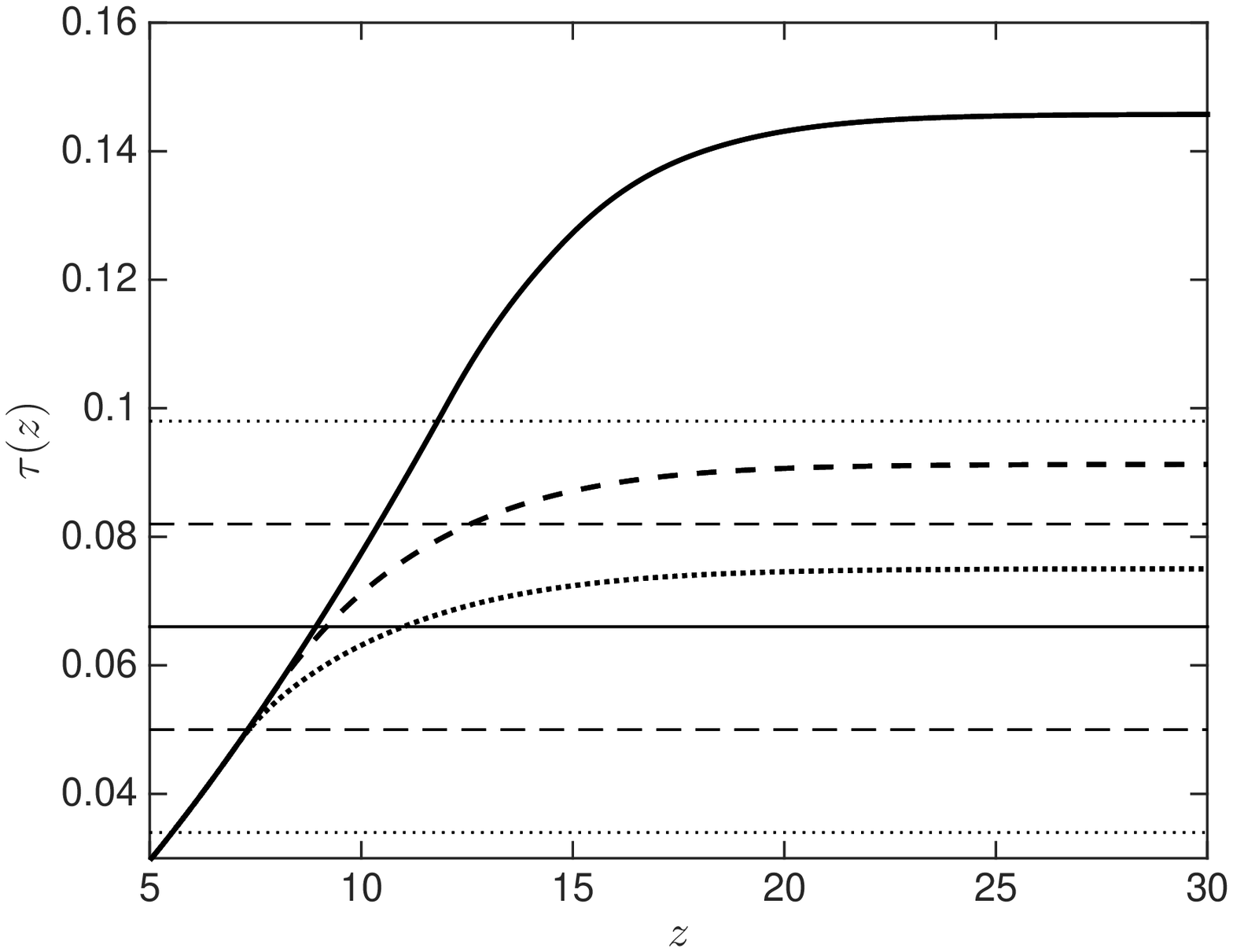}
\includegraphics[width=88mm]{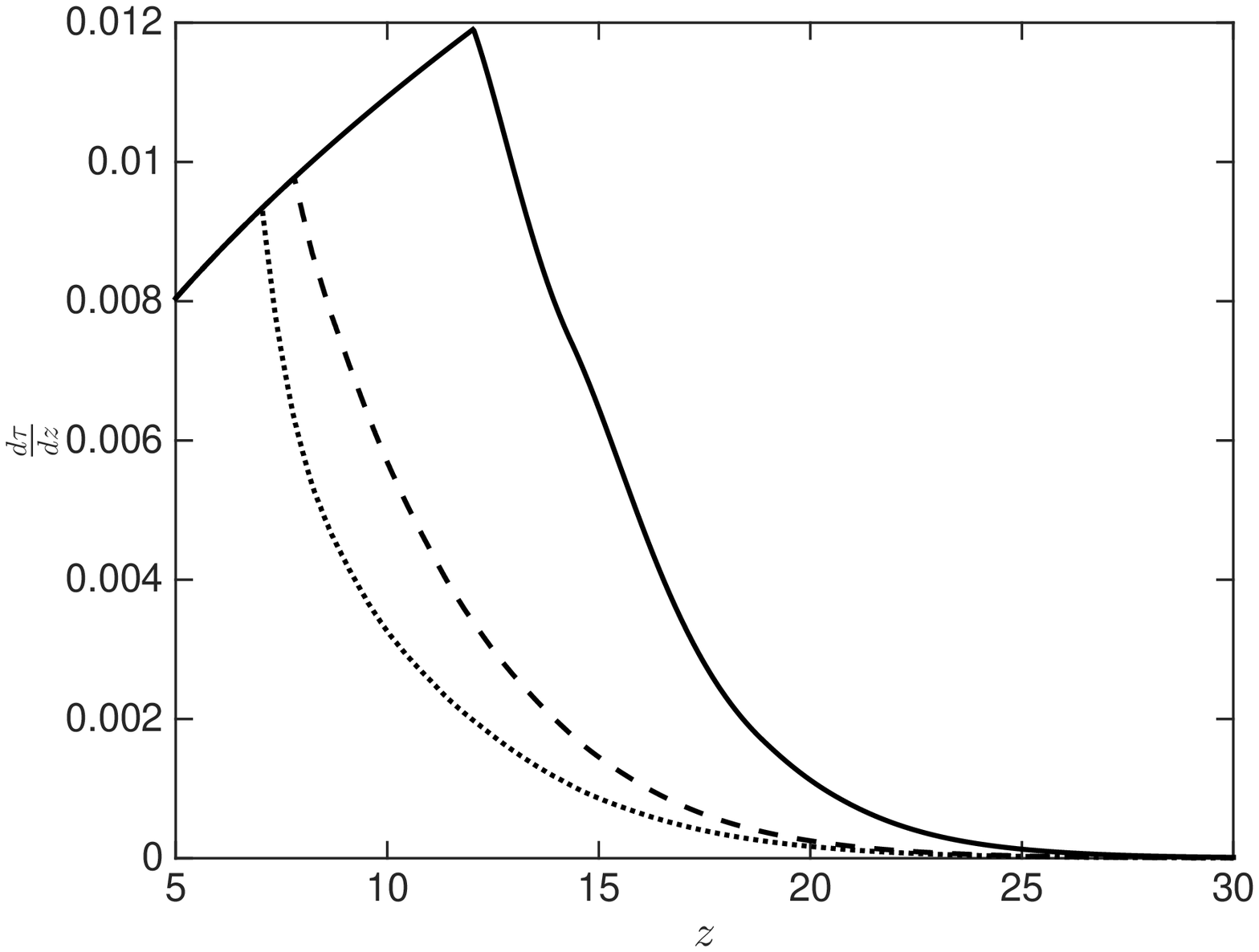}
\caption{\label{bh_plot} The comoving black hole density from remnants of Pop III stars, the fraction of the IGM ionised by X-rays emitted due to black hole accretion, and the corresponding $\tau$ and $\frac{d\tau}{dz}$. Note that the ionisation includes no contribution from stars, while $\tau$ and $\frac{d\tau}{dz}$ assume a total ionisation of $Q+Q_{\rm X}$ (with a max value of 1). The details of the calculation are explained in \S~4. The solid, dashed, and dotted curves denote $\epsilon_{\rm BH, duty}=1$, 0.1, and 0.001, respectively.   The horizontal solid, dashed, and dotted lines in the $\tau$ panel denote the \emph{Planck} measurement, and the 1$\sigma$ and 2$\sigma$ errors. Black hole growth in the  $\epsilon_{\rm BH, duty}=1$ case is stopped by hand at $z\approx 11$ in order to avoid overproducing $\rho_{\rm BH}$. This causes the IGM to begin recombining at $z<11$ - while somewhat ad-hoc, this is conservative since it limits the BH's contribution to $\tau$ (see text). }
\end{figure*}

In Figure \ref{bh_plot}, we plot $\rho_{\rm BH}(z)$ and $Q_{\rm X}(z)$ for $\epsilon_{\rm BH, duty}=1,~ 0.1$ and 0.01. It is clear that for a duty cycle of 1 the ionisation is inconsistent with \emph{Planck}. The optical depth from $Q_{\rm X}$ alone is $\tau = 0.122$, and taking the total ionisation as the sum $Q+Q_{\rm X}$ (with a max of 1) yields $\tau=0.15$.  For $\epsilon_{\rm BH, duty}=0.1$ and 0.01, taking the total ionisation as $Q+Q_{\rm X}$ gives $\tau=0.091$ and $\tau=0.075$, respectively. Thus, only for  $\epsilon_{\rm BH, duty} \lesssim 0.01$ is this model compatible with the 1$\sigma$ \emph{Planck} limits.

\section{Discussion and Conclusions}
We have performed analytic calculations of reionisation including separate contributions from Pop II stars in atomic cooling haloes and  Pop III stars in minihaloes with and without LW feedback. For haloes above the atomic cooling threshold we considered two different models, one with a constant star formation efficiency and the other with a constant SFR-halo mass relation leading to a redshift-dependent star formation efficiency. We calibrate both of these models with the  observed $z\approx 6$ UV LF.  We also incorporated a simple treatment of metal enrichment in minihaloes due to Pop III supernovae.

Without LW feedback or metal enrichment, a minihalo star formation efficiency greater than $\sim$a few$\times10^{-4}$ creates an early partial reionisation incompatible with the \emph{Planck} optical depth (assuming $f_{\rm esc, m} = 0.5$). In a $10^6 M_\odot$ minihalo, this star formation efficiency corresponds to $\sim 50 M_\odot$ of stellar mass. Thus, typical minihaloes could not form more massive Pop III stars. To demonstrate the robustness of this conclusion, we vary each model parameter, aside from those associated with minihaloes, and find that our conclusions do not change significantly. We note that the ionising efficiency of atomic cooling haloes cannot be lowered significantly from the fiducial value or reionisation will not be complete by $z=6$, inconsistent with measurements of the Ly$\alpha$ forest. With our empirical calibration of $f_{\rm *, a}$ and $\eta_{\rm ion,a}=4000$, the escape fraction cannot be less than $f_{\rm esc,a}\sim 0.1$ or reionisation will occur too late.

When we include our self-consistent treatment of LW feedback and/or metal enrichment we find that the star formation efficiency in (the massive, $M>M_m$) minihaloes can be much higher without violating the \emph{Planck} constraints: $f_{\rm *,m}\sim {\rm a ~few}\times10^{-3}$. This leads us to the main conclusion of our paper. Without LW feedback as strong as the model used in our paper or metal enrichment, massive Pop III stars in minihaloes will lead to an optical depth of the CMB that is inconsistent with \emph{Planck} observations. Another important conclusion of our paper is that the limit on the total density of large Pop III stars formed over cosmic time due to the 1-$\sigma$ \emph{Planck} constraints is roughly in the range of $10^4-10^5~M_\odot~{\rm Mpc^{-3}}$, irrespective of the feedback prescription (i.e. LW, LW+metals, etc). For reference, we note that this is $1.6 \times10^{-6} - 1.6\times10^{-5}$ of the total baryon density. We also point out that this fraction does not correspond to the density of Pop III remnants remaining today because most of the stars have lifetimes much shorter than the age of the Universe.

For our cases without LW feedback or metal enrichment although we quote constraints in terms of $f_{\rm *, m}$, the relevant quantity is 
$\epsilon_{\rm m} = f_{\rm *, m} \eta_{\rm ion,m} f_{\rm esc, m}$.  Thus, if the escape fraction were lower than our assumed $f_{\rm esc,m}=0.5$ by some factor, the corresponding limit on $f_{\rm *,m}$ would go up by the same factor.

Note that in our model we have ignored the baryon-dark matter streaming velocity \citep{2010PhRvD..82h3520T}. At high redshifts this effect can reduce the efficiency of star formation in minihaloes. However, since most of the contribution to $\tau$ in our model comes from $z<20$, this effect would only reduce the star formation efficiency by a factor of a few at most \citep{2012MNRAS.424.1335F}. 

We have also considered how X-rays emitted due to the accretion of black hole remnants from Pop III stars would impact the IGM. We performed a simple estimate of X-ray ionisation from black holes produced by our fiducial model with $f_{\rm *,m}=0.001$ and found that unless the duty cycle of black hole accretion is $\epsilon_{\rm BH, duty} \lesssim 0.01$, early ionisation produces an optical depth greater than the \emph{Planck} 1$\sigma$ limits. While we emphasise that our rough estimate is somewhat model dependent (e.g. a very hard X-ray spectrum could lead to free-streaming of X-rays and weaker constraints), the result is intriguing. To form the first super massive black holes (SMBHs), which are more massive than $10^9M_\odot$ at $z\approx6$, would require that Pop III remnants grow at the Eddington limit with a duty cycle of nearly unity. This suggests that either some type of feedback \citep[e.g.][]{2012MNRAS.425.2974T} acts on most, but not all Pop III remnants if they are the seeds of the first SMBHs or that SMBHs are seeded by a different mechanism such as direct collapse black holes \citep[e.g.][]{2003ApJ...596...34B,2010MNRAS.409.1022V,2014MNRAS.442.2036D, 2014MNRAS.445.1056V, 2015arXiv150400676I}.


\section*{Acknowledgements}
We thank the anonymous referee for his/her thorough review and useful comments. EV was supported by the Columbia Prize Postdoctoral Fellowship in the Natural Sciences.  ZH was supported by NASA grant NNX11AE05G.  GLB was supported by National Science Foundation grant 1008134 and NASA grant NNX12AH41G.

\bibliography{paper}

\end{document}